\documentclass{article}
\usepackage{arxiv}
\usepackage[utf8]{inputenc} 
\usepackage[T1]{fontenc}    
\usepackage{hyperref}       
\usepackage{url}            
\usepackage{booktabs}       
\usepackage{amsfonts}       
\usepackage{nicefrac}       
\usepackage{microtype}      
\usepackage{lipsum}		
\usepackage{graphicx}
\usepackage{natbib}
\usepackage{doi}

\usepackage[space]{grffile}
\usepackage{latexsym}
\usepackage{textcomp}
\usepackage{longtable}
\usepackage{tabulary}
\usepackage{booktabs,array,multirow}
\usepackage{amsfonts,amsmath,amssymb}
\hypersetup{colorlinks=false,pdfborder={0 0 0}}
\usepackage{etoolbox}

\usepackage{tikz}
\usepackage{float}
\usepackage{enumitem}
\usepackage{caption}
\hypersetup{
    colorlinks=true,
    linkcolor=blue,
    citecolor=blue,
    filecolor=magenta,      
    urlcolor=cyan,
    }
\newtheorem{theorem}{Theorem}

\usepackage{subcaption}
\usepackage{xr}
\newcommand{\argmin}{\normalfont\text{argmin}}

\usepackage[utf8]{inputenc}
\usepackage[english]{babel}

\usepackage{siunitx}

\title{Domain Adaptation under MNAR Missingness}

\renewcommand{\shorttitle}{Domain Adaptation Under MNAR Missingness}
\author{Tyrel Stokes \\ 
Department of Population Health \\
NYU Grossman School of Medicine \\
\AND
Hyungrok Do \\
Department of Population Health \\
NYU Grossman School of Medicine \\
\AND
Saul Blecker \\
Departments of Population Health and Medicine \\
NYU Grossman School of Medicine \\
\AND
Rumi Chunara \\
Department of Biostatistics, NYU School of Public Health \\
Department of Computer Science, NYU Tandon School of Engineering \\
\AND
Samrachana Adhikari \\
Department of Population Health\\
NYU Grossman School of Medicine \\
}

\date{}
\begin{document}
\newcommand{\X}{\boldsymbol{X}}
\newcommand{\Y}{\boldsymbol{Y}}
\newcommand{\M}{\boldsymbol{R}}
\newcommand{\Z}{\boldsymbol{Z}}
\newcommand{\R}{\boldsymbol{R}}
\newcommand{\z}{\boldsymbol{z}}
\newcommand{\x}{\boldsymbol{x}}
\newcommand{\Xo}{\X_o}
\newcommand{\Xu}{\X_u}
\newcommand{\xu}{\x_u}
\newcommand{\tx}{\X^\star}
\newcommand{\txu}{\tx_u}
\newcommand{\indep}{\perp \!\!\! \perp}
\newcommand{\one}{\boldsymbol{1}}
\newcommand{\pto}{\overset{p}{\to}}
\newcommand{\plim}{plim\,}
\newcommand{\plimn}{\underset{n\to\infty}{\plim}}
\newcommand{\hx}{\hat{\X}}
\newcommand{\hxu}{\hx_u}
\newcommand{\bp}{\mathbb{P}}
\newcommand{\bps}{\bp^{S}}
\newcommand{\bpt}{\bp^{T}}
\newcommand{\mpt}{\bp^{T}}
\newcommand{\mps}{\bp^{S}}
\newcommand{\E}{\mathbb{E}}
\newcommand{\Eps}{\E_{\bps}}
\newcommand{\Ept}{\E_{\bpt}}
\newcommand{\for}{\normalfont \text{for }}

\maketitle
\begin{abstract}
Current domain adaptation methods under missingness shift are restricted to Missing At Random (MAR) missingness mechanisms. However, in many real-world examples, the MAR assumption may be too restrictive. When covariates are Missing Not At Random (MNAR) in both source and target data, the common covariate shift solutions, including importance weighting, are not directly applicable. We show that under reasonable assumptions, the problem of MNAR missingness shift can be reduced to an imputation problem. This allows us to leverage recent methodological developments in both the traditional statistics and machine/deep-learning literature for MNAR imputation to develop a novel domain adaptation procedure for MNAR missingness shift. We further show that our proposed procedure can be extended to handle simultaneous MNAR missingness and covariate shifts. We apply our procedure to Electronic Health Record (EHR) data from two hospitals in south and northeast regions of the US. In this setting we expect different hospital networks and regions to serve different populations and to have different procedures, practices, and software for inputting and recording data, causing simultaneous missingness and covariate shifts. 

\end{abstract}%

\section{Introduction}

A central goal in many modern statistical learning applications is to develop models that can be generalized to unseen data sources. For example, consider the context of electronic health record (EHR) data across different hospital networks, where we want to develop a prediction model using clinical information from the EHR. Due to medical privacy concerns, resource constraints, and regulations, we are often unable to combine data from different networks. Instead, we may train the model on a source hospital and adapt it to make predictions on target hospitals outside the network. This problem is formally an example of domain adaptation in statistical learning \citep{ben2010theory}. In general, we might not expect the underlying probability distributions from different data sources to be equivalent and thus the domain adaptation literature is concerned with building robust procedures for handling shifts in the distribution of the underlying data \citep{ben2006analysis,ben2010theory,gretton2009covariate,zhang2013domain}. In particular, the domain adaptation literature has concerned itself with three different types of distributional shifts between the source and target domains \citep{nair2019covariate}. These include, the covariate shift, the concept shift, and the prior probability or label shift. The covariate shift refers to the situations when distributions of covariates $X$ are different between the source ($S$) and target ($T$) domains, i.e. $\bp^{T}(\X) \neq \bp^{S}(\X)$.  In the concept shift, the source and target domains have different conditional distributions of the outcome $Y$, i.e. $\bp^{T}(\Y|\X) \neq \bp^{S}(\Y|\X)$.  Finally in the label shift, the marginal outcome distributions are different but the covariates given the outcome remain the same, $\bp^{T}(\Y) \neq \bp^{S}(\Y)$ and $\bp^{T}(\X|\Y) = \bp^{S}(\X|\Y)$. See \citet{nair2019covariate} for a survey on covariate shift domain adaptation, \citet{gama2014survey} for a survey on concept shift, and \citet{garg2020unified} for a unified approach on label shift.

In the context of the EHR data, there are often additional types of shifts which can impact our inferences and predictions. Missing data, for example, are prevalent in the EHR data and may fall into one of the three categories of missing mechanisms \citet{rubin1976inference}. The first and simplest is called missing completely at random (MCAR) which is when the probability of missingness is independent of both observed and unobserved data. The second is missing at random (MAR), where the probability of missingness depends only on the observed data. And finally, all other data mechanisms where the probability of missingness may depend on unobservables is called missing not at random (MNAR). The extent or nature of the missingness can depend on the data source leading to missingness shift between the source and target data. Similar to covariate shift, missingness shift then refers to the case when the distribution of missing values is different in the source and target data, for example $\bp^{T}(\M|\X)\neq \bp^{S}(\M|\X)$, with $\M$ denoting an indicator variable taking 1 when the variable is observed (i.e. not missing) and 0 otherwise.

Missingness shift in the context of domain adaptation remains underexplored. In fact, this topic has only recently been explored in one paper by \cite{zhou2023domain} for MAR mechanism. However, in many real-world contexts, MAR mechanisms can be too restrictive of an assumption. In EHR data, unobserved patient characteristics can influence the missingness of key variables leading to MNAR mechanism. For example, the likelihood of a phsycian ordering a test may depend on symptoms which may not be recorded in the EHR data. This information is unobserved with respect to the missingness prediction task. Additionally, there may be complex interactions between hospital network policy, procedure, and culture of record keeping and the patient or encounter itself which can influence the missingness of particular variables. Thus the missingness of important clinical variables can depend on both observed and unobserved factors, implying MNAR mechanism. Further, we expect these MNAR missingness patterns to vary across data sources since there may be differences in methods and procedures for EHR data collection and management. This motivates our study of missingness shift under MNAR mechanisms.

\begin{figure}
    \centering
    \includegraphics[scale=.5]{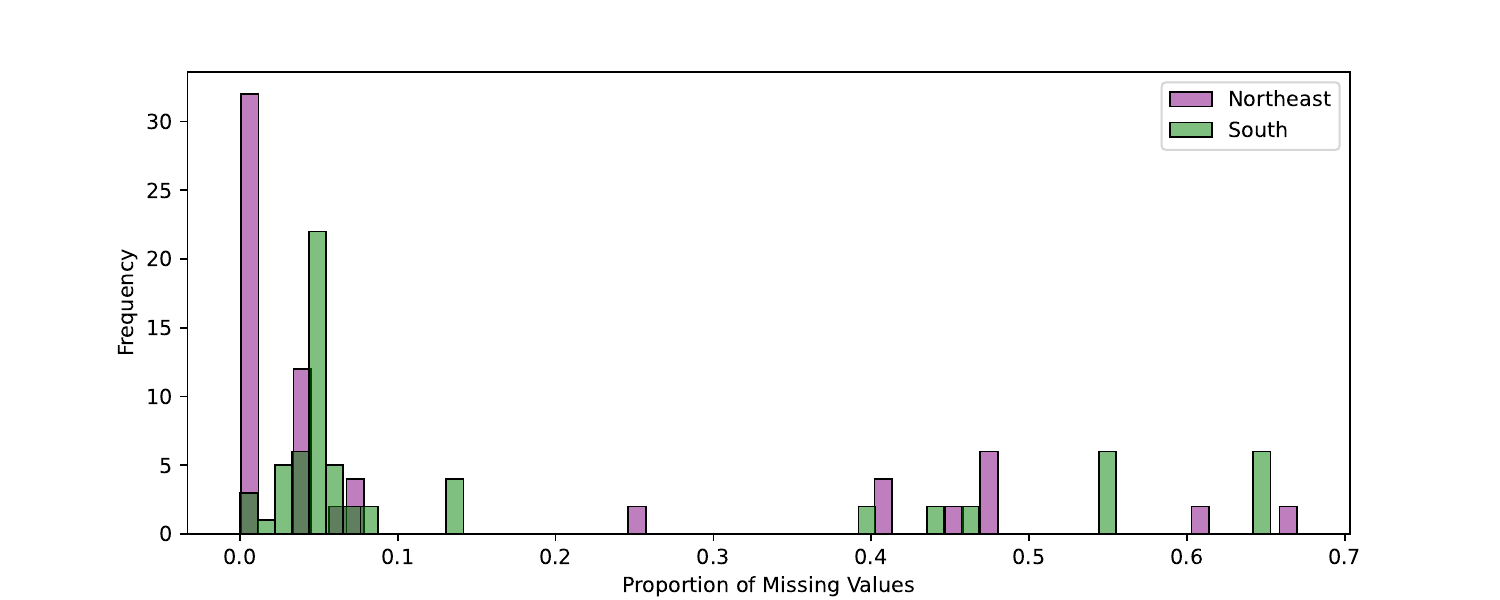}
    \caption{A histogram displaying distribution of the percent of missingness for 57 covariates in  hospitals from two different regions, the South and Northeast, from the Philips eICU dataset.}
    \label{fig:miss-dist}
\end{figure}

As an illustration, we summarize missingness in 57 covariates from the electronic intensive care units (eICU) data from the Philips eICU program \cite{pollard2018eicu} for hospitals in two US regions, South and Northeast in Figure \ref{fig:miss-dist}. We discuss the application in more detail in Section \ref{sc:application}. The first thing to note in the Figure is the difference in average missingness across all covariates, shown by the two vertical lines. In the Southern region, the overall average missingness is nearly 18\% and in the Northeast it is approximately 13\%, which is about 5\% difference. Looking more closely at the displayed distribution, we can see that one of the largest differences is the number of covariates with no missingness at all. In the Northeast region, more than 30 covariates have no missingness whereas in the Southern region this number is fewer than 5. Such stark difference is likely to be at least partly a result of policy or procedural differences across the geographic regions, highlighting the many ways in which missingness may be different across different hospital networks in EHR data. Therefore, when adapting a model across hospital networks, care is required to ensure that we are able to properly account for the differences in missingness distribution between the source and target in addition to the covariate shift traditionally explored in the domain adaptation literature. 

In this work, we contribute to the domain adaptation literature in three main ways. First, we show that under MNAR missingness shift, the importance weighting strategy proposed by \cite{zhou2023domain} for MAR missingness shift does not directly apply since the conditional outcome distribution will be shifted under shifted missingness. We further show that a more general version to the reweighting approach can still apply whereby we first impute the missingness data in both the source and the target, separately. Second, MNAR missingness imputation is a difficult problem with an emerging literature - both in traditional statistics and machine learning fields - to identify MNAR missingness structures under weak parametric assumptions. For these different imputation methods, we discuss various kinds of MNAR structures that can be identified and the conditions necessary for identification. Finally, using an extensive simulation study we show that when one has access to an imputation procedure capable of recovering the MNAR structures in both the source and target domains, our proposed domain adaptation procedure can additionally handle covariate shift using standard covariate shift techniques such as importance reweighting.
 
 We outline the problem and show the conditions under which missingness shift can be reduced to an imputation problem in Section \ref{sc:prob}. In Section \ref{sc:impute} we discuss a selection of imputation procedures suitable for MNAR mechanisms, and test them in simulation experiments in Section \ref{sc:sim}. Finally, we apply our novel procedure for handling MNAR missingness shift and covariate shift simultaneously to a real-world EHR data example in Section \ref{sc:application}.

\par\null

\section{Domain Adaptation with Missing Not At Random Shift}\label{sc:prob}

\subsection{Notation and Setting}
Suppose there are two domains, a source domain for which we have access to the observed covariates and outcome for the full sample, and a target domain for which we only have access to a sub-sample of the observed covariate data and no outcome data. Our goal is to minimize the prediction error in the target domain. In our set-up we additionally assume that there may be missing covariates and that the missingness mechanism may be different in the source and target domains.

 First, we introduce the notation for missingness under a counterfactual framework. Consider the following partition of the covariate space, $\X =[\Xo,\Xu]$, where $\Xo$ represent columns of covariates which are not missing and $\Xu$ represent the counterfactual of covariates with missingness (i.e., the true underlying values). The factually observed variables are denoted $\txu$ and defined in the case that $\Xu$ is a single-dimensional vector as:
\begin{align}\label{eqn:factual}
    \txu = \begin{cases}
\Xu, & \text{if} ~ \M =1\\
\text{NA},  & \text{if} ~ \M =0
\end{cases},
\end{align}
where $\M$ is the indicator variable taking 1 when the variable is not missing and 0 otherwise. We assume throughout that there is no missingness in the outcome. In general, $\M$ may be a matrix of indicators over many covariates. In this case, the matrix $\txu$ with the same dimension as $\Xu$ is defined analogously through its elements $\txu(i,j)$ which take values $\Xu(i,j)$ when $\M_{ij}$ is 1 and NA otherwise.

Let $\mathbb{P}^S$ and $\mathbb{P}^T$ denote data distributions from the source and target domains, respectively. The domain adaptation is characterized by the goal to find the best predictor, which has the smallest expected loss, under $\mathbb{P}^T$ when we have observed samples from $\mathbb{P}^S(\Y,\Xo,\txu,\M)$ and a (sub)-sample from $\mathbb{P}^T(\Xo,\txu,\M)$. When the function class $\mathcal{G}$ is unrestricted and the loss is $\mathcal{L}_2$, the domain adaptation goal is formulated as follows:
\begin{align}
    \underset{g\in \mathcal{G}}{\argmin}&~\mathbb{E}_{\mathbb{P}^{T}}[(\Y - g(\Xo,\txu,\M))^2]\nonumber\\
    &= \mathbb{E}_{\mathbb{P}^{T}}[\Y|\Xo,\txu,\M].
\end{align}
Above we assume that $\mathbb{P}^T(\Y|\Xo,\Xu,\M) = \mathbb{P}^S(\Y|\Xo,\Xu,\M)$ and that $\mathbb{P}^T(\Xo,\Xu) = \mathbb{P}^S(\Xo,\Xu)$, but that the missingness mechanism may be different across domains, i.e., $\mathbb{P}^T(\M|\Xo,\Xu) \neq \mathbb{P}^S(\M|\Xo,\Xu)$.

\subsection{Importance Sampling Weights to Account for Covariate Shift}\label{sc:related}
Analyzing the adaptation of prediction models under distributional shift is an active area of research \citet{nair2019covariate, gama2014survey, garg2020unified}. As discussed in the previous section there are three types of distributional shifts studied in the domain adaptation literature - covariate shift, concept shift, and label shift. As we will show, our work most predominantly shares features with a subset of the covariate shift domain adaptation problem, where we additionally assume that the conditional outcome distributions are equal in the source and target domains, $\bp^T(\Y|\R,\Xo,\Xu) = \bp^S(\Y|\R,\Xo,\Xu)$. The most common approach to remove bias in outcome prediction in target data using the model from the source data due to this type of covariate shift, assuming no other shifts are present, is to estimate importance sampling weights $w(x) = \bp^{T}(x)/\bp^{S}(x)$ and re-weight the source data accordingly. In fact, for a general loss function $l(\cdot)$ and a regression function $g(X)$, the importance sampling weights transform the domain adaptation problem into an expectation with respect to the source data as shown below.
\begin{align}
\underset{g\in \mathcal{G}}{\argmin}~\mathbb{E}_{\mathbb{P}^{T}}[l(y,g(x)] &= \underset{g\in \mathcal{G}}{\argmin}\iint l(y,g(x))\bp^T(y|x)\bp^T(x)dydx\\
&= \underset{g\in \mathcal{G}}{\argmin}\iint l(y,g(x))\bp^S(y|x)\frac{\bp^T(x)}{\bp^S(x)}\bp^S(x)dydx\\
&= \underset{g\in \mathcal{G}}{\argmin} ~ \mathbb{E}_{\mathbb{P}^{S}} \left[\frac{\bp^T(\X)}{\bp^S(\X)} l(\Y,g(\X)) \right],\label{eq:minimizer}
\end{align}
whenever $\bp^T(\X) > 0$, $\bp^S(\X) > 0$, i.e., $\bp^{T}$ is absolutely continuous with respect to $\bp^{S}$, or equivalently, the support of the target and source domains have overlap. 

\begin{figure}
    \centering
    \includegraphics[scale = .4]{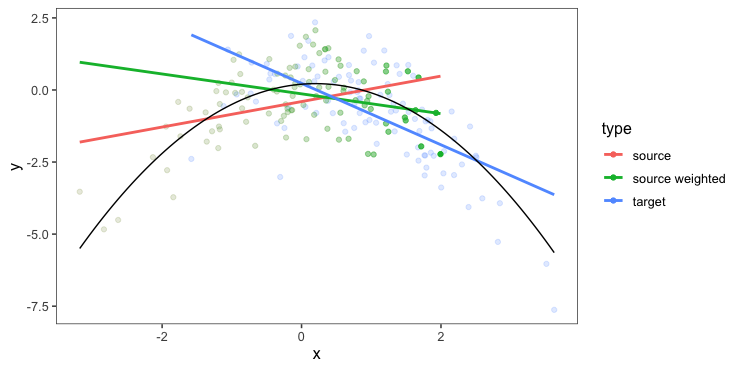}
    \caption{Finite sample simulation showing how importance weights correct the source approximation towards the target approximation even under model mispecification. In this simulation we assume that the source and the target have the same conditional outcome distribution (black curve), but different covariate distributions in a classic covariate shift set-up. Red line shows linear approximation using source data alone, blue line shows linear approximation using target data alone, and green line is the approximaiton using source data with importance sampling weights to account for covairate shift between source (green dots) and target points (blue dots).}
    \label{fig:weight-simulation}
\end{figure}

Importance sampling weights work by up-weighting samples from the source data which are more similar to collections of covariates found in the target data. As an illustration, the phenomenon of importance sampling for a finite sample can be visualized in Figure \ref{fig:weight-simulation}, where we simulated data for both the source and the target from the same conditional outcome distribution: ($\Y \sim N(\alpha + x\beta_1 + x^2\beta_2, \sigma), \alpha = 0.2, \beta = (0.2,-0.5),\sigma = 1$). The black curve shows the true underlying conditional mean function. However, the two data sets differ in their covariate distributions with ($x_{S} \sim N(0,1)$ for the source (green dots) and $x_T \sim N(1,1)$) for the target (blue dots). The opaque dark green dots represent the covariate space of the source that is most closely aligned with that of the target and for which the importance weights are larger. Therefore, source data most rightward in its covariate space sitting in the bulk of the blue target data gets the most weight, while source points to the far left are given very little weight.  

Next, we used a simple linear approximation model to fit the data separately for the source data (red line), target data (blue line), and source data with importance sampling weights (green line). Because most of the source data are concentrated towards the left of the true conditional mean curve, the linear approximation for the source data has a positive slope. Similarly, the linear approximation using the target data only (blue dots) has a negative slope. In such a case, a direct transfer of the source model to the target data without any adjustment will perform poorly in predicting outcome in the target. Finally, the linear approximation using the importance sampling weights (green line) is much closer to the target approximation (blue line) and the underlying conditional mean function in the target covariate space (black line). Therefore, through domain adaptation, we are able to closely align the approximation using the source data to the true underlying curve in the covariate space for the target data. 


However, the re-weighting strategy with importance sampling weights for covariate shift is not directly applicable to MNAR missingness shift without any additional assumption. This is because when the MNAR missingness mechanisms vary across the domains, the observed data distributions will be different even if the conditional outcome distributions are equivalent. That is, even when $\bpt(\Y|\Xo,\Xu,\M) = \bp^S(\Y|\Xo,\Xu,\M)$, if $\bp^T(\M|\Xo,\Xu)$ $\neq$ $\bp^S(\M|\Xo,\Xu)$ then $\bp^T(\Y|\Xo,\txu,\M)$ $\neq$ $\bp^S(\Y|\Xo,\txu,\M)$. We show the proof of this claim in the Supplementary Materials. Note that, the conditional distributions $\bpt(\Y|\Xo,\Xu,\M)$ and $\bp^S(\Y|\Xo,\Xu,\M)$ are functions of both factual and counterfactual $X$ as well as observation pattern $\M$, while $\bp^T(\Y|\Xo,\txu,\M)$ and $\bp^S(\Y|\Xo,\txu,\M)$ are functions of observed data only. Therefore, even if we were able to estimate the importance weights $(\frac{\bpt(\Xo,\txu,\R)}{\bps(\Xo,\txu,\R)})$ using the observed data, we would not simply be able to re-weight the samples in the source domain without also adapting the outcome model or covariates in some way. We discuss our proposed solution for such MNAR shifts next.

\subsection{Domain Adaptation with MNAR Shift as an Imputation Problem}\label{sc:reduction}

In this section, we show that by combining modern MNAR missingness imputation techniques (which we discuss in next section) and classic tools from the domain adaptation literature, such as re-weighting with importance sampling weights, we are able to extend missingness shift to a subset of MNAR structures. In particular, in Theorem \ref{thm:1} we show that when the conditional mean is a continuous function, the problem of domain adaptation under MNAR missingness shift can be reduced to imputing the missing data in both the source and the target separately.

\begin{theorem}{}\label{thm:1}
    Suppose that we have access to an imputation procedure $\hxu$ as well as a non-parametric estimator of the conditional mean, $\hat{\mu}(\cdot)$, such that $\hat{\mu}(\Y|\Xo,\Xu,\M) \pto \E_{\bp}[\Y|\Xo,\Xu,\M]$. In this case we can show that the asymptotic reducible error with respect to the mean-squared error (MSE) is a function of the imputation error $\mathbf{e} = \hxu - \Xu$. Specifically the non-irreducible error for the adaptaion problem can be expressed as  
\begin{align*}
&\Ept[(\mathbb{E}_{\mathbb{P}^{T}}[\Y|\Xo,\txu,\M] - \hat{\mu}(\Y|\Xo,\txu = \hat{\X}_u,\M))^2]\\
& \pto 
 ~ P(\M =0)\mathbb{E}_{\mathbb{P}^{T}}[(\mathbb{E}_{\mathbb{P}^{T}}[\Y|\Xo,\txu,\M=0] \nonumber\\
& \qquad - \plimn\hat{\mu}(\Y|\Xo,\txu= \Xu + \mathbf{e},\M=0))^2|\M=0],
\end{align*}
\end{theorem}

The proof can be seen in the Supplementary materials. In the above expression, we can see that the non-irreducible error goes to zero asymptotically if either the probability of missingness ($\mathbb{P}(\M=0)$) goes to 0 or the imputation estimate converges in probability to the counterfactual data ($\hat{\X}_u\pto \Xu$). If the dimension of $\Xu$ is large, it may be unrealistic for $\mathbb{P}(\M=0)$ to be exceedingly small. However, in the case that the true conditional expectation function $\mathbb{E}_{\mathbb{P}^{T}}[\Y|\Xo,\Xu,\M]$ is continuous in $\Xu$, the asymptotic non-irreducible error is bounded by the imputation error $\mathbf{e} = \hat{\X}_u-\Xu$. In this sense, the problem reduces to achieving the lowest imputation error possible. In practice, this error will depend on both the imputation error on the source and target data, even though we only see the target imputation error appear directly in the above expression. The imputation error in the source domain will bound the accuracy of learning the correct conditional outcome distribution. 

\begin{figure}
\centering
\scalebox{.8}{
\begin{tikzpicture}
\node (xo) at (-1,1) {$\Xo$};
\node (xu) at (-1,-1) {$\Xu$};
\node (m) at (0,0) {$\M$};
\node (y) at (2,0) {$\Y$};
\node (txu) at (1,-1) {$\txu$};
\draw[->] (xo) to (m);
\draw[->] (xu) to (m);
\draw[->] (xu) to (txu);
\draw[->] (m) to (txu);
\draw[->] (xo) to (y);
\draw[->] (xu) to[out =30,in = -150] (y);
\draw[<->] (xu) to (xo);
\end{tikzpicture}
}
\caption{An example of missing data DAG (mDAG). $\Xo$ is the covariates with no missing, $\Xu$ is the counterfactual of covariates with missingness, $\Xu^{\star}$ represents the factually observed variables with missingness, and $\M$ is the missingness indicator. $\Y$ is the outcome variable.}
\label{fig:mdag-1}
\end{figure}
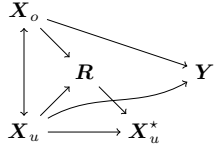

Therefore, under simplifying sets of assumptions it is possible to reduce the dependence of the generalization error to the target imputation procedure only. While, we do not outline the universe of these assumptions here, we provide an example for intuition. For example, under the assumption that there is no direct cause from the missingness indicator to the outcome, as represented in the missing data Directed Acyclic Graph in Figure \ref{fig:mdag-1}, it is possible to learn the conditional outcome distribution with only the complete case data (proof in the Supplementary Materials).

In most cases, however, one might expect there to be both traditional covariate shift and missingness shift. In the Supplementary Materials we show that when one has access to an imputation procedure which converges in probability to the counter-factual missing variables for both the source and target then simultaneous covariate and missingness shift can be tackled using the importance weighting approach. Specifically, one first imputes the missing data in the source and target and then estimates the importance weights using the imputed values. The relevant importance weights are $\bpt(\Xo,\Xu,\R)/\bps(\Xo,\Xu,\R)$. In practice, it is often simpler to transform this into a classification problem by estimating $\bp(\text{Sample} \in T|\Xo,\Xu,\R) / \bp(\text{Sample} \in S|\Xo,\Xu,\R)$ which is proportional to the original weights. Once the estimated weights are obtained, one can estimate the target conditional outcome by using samples in the source data re-weighted by the importance weights. We demonstrate this approach in Section \ref{sc:application}. Notice that a correct imputation device for both the target and source is required when there is covariate shift present in addition to the missingness shift.  In the next section, we review several traditional and machine learning approaches to impute MNAR data.

\section{Imputation Methods and Identification}\label{sc:impute}

As we show in the previous section, finding a solution to the MNAR missingness shift problem reduces to imputation problem for MNAR mechanisms under a set of identifying assumptions. We extend missingness shift to handle MNAR mechanism by leveraging existing work in traditional statistics and machine learning for imputing MNAR missingness structures. 
Traditional statistics approaches to imputing MNAR data, including the Bayesian latent variable approach, largely reduce to making parametric assumptions about the unobserved data for identifiability (see \cite{Little2019} for a survey of traditional statistics approaches). More recently a non-parametric identification literature has developed to establish sufficient conditions for identification of the complete-data distribution as well as the entire missingness mechanism $\bp(\M|\X)$. Much of this literature starts with missing data Directed Acyclic Graphs (DAGs) or mDAGs \citep{Mohan2013} which are graphical models to represent the joint distribution over observed variables $\Y, \Xo, \Xu^\star$, the missingness indicator $\R$ and the counterfactual complete variables we would have liked to observe had there been no missingness $\Xu$. Figure \ref{fig:mdag-1} (discussed above) is an example of an mDAG. Non-parametric identification approaches of the complete-data distribution largely proceed by non-parametric identification of the propensity scores \citep{Bhattacharya2020} or the odds-ratio \citep{Nabi2020}, the latter of which is necessary and sufficient for non-parametric identification. Within this approach, however, it can be proven that some MNAR structures of interest cannot be identified, including self-censoring where the missingness of a variable is caused by its unobserved value itself, as shown by $\Xu \longrightarrow \M$ edge in Figure \ref{fig:mdag-1}. A simple example would be in an income survey where especially high or low true incomes might cause non-respone. The complete-data distribution under this missingness structure can only be identified under parametric assumptions. Simultaneously, machine learning MNAR literature \citep{Ipsen2020,Ma2021} has independently focused on relaxing the parametric restrictions necessary for identification while handling more graphical structures such as self-censoring, although there is minimal overlap with the mDAG literature. 

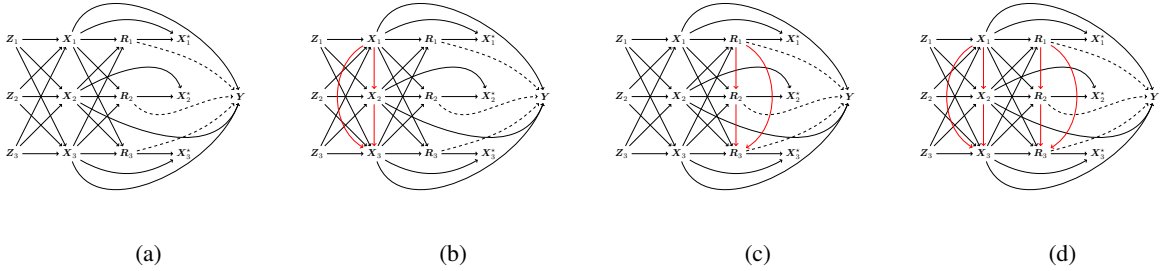
\begin{figure}[]
\centering
\begin{subfigure}{.24\textwidth}
\scalebox{.38}{
 \begin{tikzpicture}
\node (z1) at (-2,2) {$\Z_1$};
\node (z2) at (-2,0) {$\Z_2$};
\node (z3) at (-2,-2) {$\Z_3$};
\node (x1) at (0,2) {$\X_1$};
\node (x2) at (0,0) {$\X_2$};
\node (x3) at (0,-2) {$\X_3$};
\node (r1) at (2,2) {$\R_1$};
\node (r2) at (2,0) {$\R_2$};
\node (r3) at (2,-2) {$\R_3$};
\node (x1s) at (4,2) {$\tx_1$};
\node (x2s) at (4,0) {$\tx_2$};
\node (x3s) at (4,-2) {$\tx_3$};
\node (y) at  (6,0) {$\Y$};
\draw[->] (z1) to (x1);
\draw[->] (z1) to (x2);
\draw[->] (z1) to (x3);
\draw[->] (z2) to (x1);
\draw[->] (z2) to (x2);
\draw[->] (z2) to (x3);
\draw[->] (z3) to (x1);
\draw[->] (z3) to (x2);
\draw[->] (z3) to (x3);
\draw[->] (x1) to (r1);
\draw[->] (x1) to (r2);
\draw[->] (x1) to (r3);
\draw[->] (x2) to (r1);
\draw[->] (x2) to (r2);
\draw[->] (x2) to (r3);
\draw[->] (x3) to (r1);
\draw[->] (x3) to (r2);
\draw[->] (x3) to (r3);
\draw[->] (r1) to (x1s);
\draw[->] (r2) to (x2s);
\draw[->] (r3) to (x3s);
\draw[->] (x1) to[out =30, in = 150] (x1s);
\draw[->] (x3) to[out =-30, in = -150] (x3s);
\draw[->] (x2) to[out =30, in = 110] (x2s);
\draw[->] (x1) to[out =70, in = 110] (y);
\draw[->] (x3) to[out =-70, in = -110] (y);
\draw[->] (x2) to[out =-30, in = -110] (y);
\draw[->,dashed] (r1) to[out = -15, in = 135] (y);
\draw[->,dashed] (r2) to[out = -45, in = 180] (y);
\draw[->,dashed] (r3) to[out = 15, in = -135] (y);
\end{tikzpicture}
 }
\caption{}\label{fig:simdag1}
 \end{subfigure}
\begin{subfigure}{.24\textwidth}
\scalebox{.38}{
\begin{tikzpicture}
\node (z1) at (-2,2) {$\Z_1$};
\node (z2) at (-2,0) {$\Z_2$};
\node (z3) at (-2,-2) {$\Z_3$};
\node (x1) at (0,2) {$\X_1$};
\node (x2) at (0,0) {$\X_2$};
\node (x3) at (0,-2) {$\X_3$};
\node (r1) at (2,2) {$\R_1$};
\node (r2) at (2,0) {$\R_2$};
\node (r3) at (2,-2) {$\R_3$};
\node (x1s) at (4,2) {$\tx_1$};
\node (x2s) at (4,0) {$\tx_2$};
\node (x3s) at (4,-2) {$\tx_3$};
\node (y) at  (6,0) {$\Y$};
\draw[->] (z1) to (x1);
\draw[->] (z1) to (x2);
\draw[->] (z1) to (x3);
\draw[->] (z2) to (x1);
\draw[->] (z2) to (x2);
\draw[->] (z2) to (x3);
\draw[->] (z3) to (x1);
\draw[->] (z3) to (x2);
\draw[->] (z3) to (x3);
\draw[->] (x1) to (r1);
\draw[->] (x1) to (r2);
\draw[->] (x1) to (r3);
\draw[->] (x2) to (r1);
\draw[->] (x2) to (r2);
\draw[->] (x2) to (r3);
\draw[->] (x3) to (r1);
\draw[->] (x3) to (r2);
\draw[->] (x3) to (r3);
\draw[->] (r1) to (x1s);
\draw[->] (r2) to (x2s);
\draw[->] (r3) to (x3s);
\draw[->] (x1) to[out =30, in = 150] (x1s);
\draw[->] (x3) to[out =-30, in = -150] (x3s);
\draw[->] (x2) to[out =30, in = 110] (x2s);
\draw[->] (x1) to[out =70, in = 110] (y);
\draw[->] (x3) to[out =-70, in = -110] (y);
\draw[->] (x2) to[out =-30, in = -110] (y);
\draw[->,red,thick] (x1) to (x2);
\draw[->,red,thick] (x2) to (x3);
\draw[->,red,thick] (x1) to[out = 210, in = 150] (x3);
\draw[->,dashed] (r1) to[out = -15, in = 135] (y);
\draw[->,dashed] (r2) to[out = -45, in = 180] (y);
\draw[->,dashed] (r3) to[out = 15, in = -135] (y);
\end{tikzpicture}
}
\caption{}\label{fig:simdag2}
\end{subfigure}
\begin{subfigure}{.24\textwidth}
\scalebox{.38}{
\begin{tikzpicture}
\node (z1) at (-2,2) {$\Z_1$};
\node (z2) at (-2,0) {$\Z_2$};
\node (z3) at (-2,-2) {$\Z_3$};
\node (x1) at (0,2) {$\X_1$};
\node (x2) at (0,0) {$\X_2$};
\node (x3) at (0,-2) {$\X_3$};
\node (r1) at (2,2) {$\R_1$};
\node (r2) at (2,0) {$\R_2$};
\node (r3) at (2,-2) {$\R_3$};
\node (x1s) at (4,2) {$\tx_1$};
\node (x2s) at (4,0) {$\tx_2$};
\node (x3s) at (4,-2) {$\tx_3$};
\node (y) at  (6,0) {$\Y$};
\draw[->] (z1) to (x1);
\draw[->] (z1) to (x2);
\draw[->] (z1) to (x3);
\draw[->] (z2) to (x1);
\draw[->] (z2) to (x2);
\draw[->] (z2) to (x3);
\draw[->] (z3) to (x1);
\draw[->] (z3) to (x2);
\draw[->] (z3) to (x3);
\draw[->] (x1) to (r1);
\draw[->] (x1) to (r2);
\draw[->] (x1) to (r3);
\draw[->] (x2) to (r1);
\draw[->] (x2) to (r2);
\draw[->] (x2) to (r3);
\draw[->] (x3) to (r1);
\draw[->] (x3) to (r2);
\draw[->] (x3) to (r3);
\draw[->] (r1) to (x1s);
\draw[->] (r2) to (x2s);
\draw[->] (r3) to (x3s);
\draw[->] (x1) to[out =30, in = 150] (x1s);
\draw[->] (x3) to[out =-30, in = -150] (x3s);
\draw[->] (x2) to[out =30, in = 110] (x2s);
\draw[->] (x1) to[out =70, in = 110] (y);
\draw[->] (x3) to[out =-70, in = -110] (y);
\draw[->] (x2) to[out =-30, in = -110] (y);
\draw[->,red,thick] (r1) to (r2);
\draw[->,red,thick] (r2) to (r3);
\draw[->,red,thick] (r1) to[out = -30, in = 30] (r3);
\draw[->,dashed] (r1) to[out = -15, in = 135] (y);
\draw[->,dashed] (r2) to[out = -45, in = 180] (y);
\draw[->,dashed] (r3) to[out = 15, in = -135] (y);
\end{tikzpicture}
}
\caption{}\label{fig:simdag3}
\end{subfigure}
\begin{subfigure}{0.24\textwidth}
\scalebox{.38}{
\begin{tikzpicture}
\node (z1) at (-2,2) {$\Z_1$};
\node (z2) at (-2,0) {$\Z_2$};
\node (z3) at (-2,-2) {$\Z_3$};
\node (x1) at (0,2) {$\X_1$};
\node (x2) at (0,0) {$\X_2$};
\node (x3) at (0,-2) {$\X_3$};
\node (r1) at (2,2) {$\R_1$};
\node (r2) at (2,0) {$\R_2$};
\node (r3) at (2,-2) {$\R_3$};
\node (x1s) at (4,2) {$\tx_1$};
\node (x2s) at (4,0) {$\tx_2$};
\node (x3s) at (4,-2) {$\tx_3$};
\node (y) at  (6,0) {$\Y$};
\draw[->] (z1) to (x1);
\draw[->] (z1) to (x2);
\draw[->] (z1) to (x3);
\draw[->] (z2) to (x1);
\draw[->] (z2) to (x2);
\draw[->] (z2) to (x3);
\draw[->] (z3) to (x1);
\draw[->] (z3) to (x2);
\draw[->] (z3) to (x3);
\draw[->] (x1) to (r1);
\draw[->] (x1) to (r2);
\draw[->] (x1) to (r3);
\draw[->] (x2) to (r1);
\draw[->] (x2) to (r2);
\draw[->] (x2) to (r3);
\draw[->] (x3) to (r1);
\draw[->] (x3) to (r2);
\draw[->] (x3) to (r3);
\draw[->] (r1) to (x1s);
\draw[->] (r2) to (x2s);
\draw[->] (r3) to (x3s);
\draw[->] (x1) to[out =30, in = 150] (x1s);
\draw[->] (x3) to[out =-30, in = -150] (x3s);
\draw[->] (x2) to[out =30, in = 110] (x2s);
\draw[->] (x1) to[out =70, in = 110] (y);
\draw[->] (x3) to[out =-70, in = -110] (y);
\draw[->] (x2) to[out =-30, in = -110] (y);
\draw[->,red,thick] (r1) to (r2);
\draw[->,red,thick] (r2) to (r3);
\draw[->,red,thick] (x1) to (x2);
\draw[->,red,thick] (x2) to (x3);
\draw[->,red,thick] (x1) to[out = 210, in = 150] (x3);
\draw[->,red,thick] (r1) to[out = -30, in = 30] (r3);
\draw[->,dashed] (r1) to[out = -15, in = 135] (y);
\draw[->,dashed] (r2) to[out = -45, in = 180] (y);
\draw[->,dashed] (r3) to[out = 15, in = -135] (y);
\end{tikzpicture}
}
\caption{}\label{fig:simdag4}
\end{subfigure}
\caption{DAGs corresponding to various simulation settings. In all scenarios, the covariates $\X$ are dependent on $\Z$, while the missing indicator $\M$ depends on $\X$, and the outcome $\Y$ depends on $\X$. The factually observed covariates $\X^{\star}$ are generated based on Equation \ref{eqn:factual}. Additionally, the following conditions hold: (a) Each latent variable is mutually independent of the others, and the same independence holds for both the covariates and missingness indicators; (b) $\X_2$ depends on $\X_1$, and $\X_3$ depends on both $\X_1$ and $\X_2$; (c) $\R_2$ depends on $\R_1$, and $\R_3$ depends on both $\R_1$ and $\R_2$; (d) a combination of conditions from (b) and (c). The dashed lines show dependence from the missingness indicators to the outcome. In our simulations we consider the cases with and without direct causal pathways from the missingness outcomes to be different simulation DAGs, i.e 8 total simulation DAGs, but represent them in this collapsed form for parsimony of presentation.}\label{fig:sim dags}
\end{figure}

In this section, we review existing imputation methods and underlying underlying mDAG structures and assumptions for three categories of approaches that are commonly used in practice for MNAR imputation. These categories of models include joint Bayesian latent variable models \cite{Du2022}, multiple imputation by chained equations (MICE) \citep{van2011mice} and machine learning imputation models using Variational Autoencoders (VAE). For each imputation method, we discuss the identification of MNAR mechanisms within the context for eight data generating DAGs as shown in Figure \ref{fig:sim dags}. (These DAGs also inform our simulation study in the next section.)


 Let $\Y$ be the outcome vector, $\Xo = \X_1$ be the fully observed variables, $\X_u = [\X_2,\X_3]$ be the potentially unobserved counter-factual variables, $\tx$ be the vector of factually observed covariates, $\R$ be the $n \times n_u$ missingness indicator matrix where $n_u$ is the number of partially unobserved variables, and $\Z$ be the $n \times n_u$ matrix of latent variables. For simplicity, we assume $n_u = 2$ in Equation \ref{eqn::DGM} and in the simulation study. We consider the class of structural models to represent the underlying data generating models corresponding to the DAGs as
\begin{equation}\label{eqn::DGM}
\begin{aligned}
\Z_i &\sim N(\alpha_{z_i}, \sigma_z), \for i = 1,2,3\\
\X_i &\sim N(\alpha_{x_i} + f_z(\Z) + f_x(\X_{1:(i-1)}), \sigma_{x}), \for i = 2,3\\
\R_i &\sim Ber(\text{expit}(\alpha_{r_i} + g_x(\X_{1:(i-1)}))), \for i = 2,3\\
\Y &\sim N(\alpha_y + h_x(\X) + h_m(\M),\sigma_y),
\end{aligned}
\end{equation}
where $f_z(\cdot),f_x(\cdot),g_x(\cdot),$ $h_x(\cdot)$, and $h_m(\cdot)$ are potentially non-linear functions and $\alpha_{(\cdot)}$ and $\sigma_{(\cdot)}$ represent intercept and noise terms respectively. In practice, these functions can be set to either linear or generated from neural network mean-functions (sigmoid or ReLU activation functions with linear final output functions). For simulations, these functions were generated with 2 layers with 2 nodes each, 3 layers with 3 nodes each, or 4 layers with 4 nodes each. 
Notice that all of the simulation DAGs feature self-censoring, for example $\X_i \rightarrow \R_i$. As discussed in Section \ref{sc:related}, this means that these structures are not identifiable non-parametrically. To identify these MNAR missingness mechanisms it is required to make additional parametric assumptions. Therefore, all imputation methods we review below additionally make parametric assumptions but vary with respect to how restrictive those assumptions are.

\subsection{Joint Bayesian Models}
We provide the details for the joint Bayesian models in the spirit of \cite{Du2022} to jointly model outcome and missing indicator. We follow the notation presented in Equation \eqref{eqn::DGM}, such that $\X_1$ is fully observed, whereas $\X_2$ and $\X_3$ are potentially unobserved counterfactual variables. The joint Bayesian models are based on the hierarchical structure and the corresponding conditional distributions presented in Equation \eqref{eqn::DGM} and under the parametric restriction that the functions $f_z(\cdot),f_x(\cdot),g_x(\cdot),$ and $h_x(\cdot)$ are strictly linear. The goal in the Bayesian model is to sample from the conditional posterior distribution of the missing entries of the covariates $\X_2$ and $\X_3$ using MCMC or variational inference. We use the posterior means $\hx_2$ and $\hx_3$ as a point estimate for the missing entries to more directly compare with the machine learning methods which do not produce distributions over the imputed variable. The priors for the parameters can be set to be weakly informative for all parameters, as normal or truncated normal distributions with standard deviations between 1 and 5. The probability models in Equation \eqref{eqn::DGM}, then, give rise to different estimators in decreasing complexity by omitting portions of the joint distribution. 

The first two group of estimators are based on full joint model specification. First, full joint model using all four equations, including a model for the latent variables $\Z$, is refered to as the `z-Model' estimator. It specifies a relationship between all observed and unobserved variables in the simulation DAGs in Figure \ref{fig:sim dags}. Second, we define the `joint model', as the Bayesian estimator which posits a hierarchical model for the covariates $\X$, missingness indicator $\R$, and outcome $\Y$. This is equivalent to the z-model without an explicit distribution over the latent variables. 

There is a further nuance with these two full Bayesian models, the z-model and joint model. These models specify joint distributions over the outcome, but this is not strictly possible in the domain adaptation setting we specify in this manuscript when the outcome is not available in the target. We thus distinguish two additional cases of the joint and z models. The first uses the target outcome information, which we call the `leak setting', and the second only specifies the joint distribution over the outcome in the source, but ignores this information in the target case which we denote the `no leak setting'. Whether or not the outcome is included in the specification can have important implications for identification. In particular, as demonstrated in \cite{d2024behind}, the outcome should be included when in the context of stochastic imputation in order to achieve unbiased imputation. In the simulation section, we show that the Joint Bayesian models behave like stochastic imputation models in the sense that inclusion of the outcome is necessary for unbiased imputation. This is despite the fact that we only use a single point estimate for imputation in this context.

 Next, we define group of estimators based on partial Bayesian models. We refer to the `No Outcome Model' as a model over only the covariates $\X$ and missingness indicators $\R$, and the `No Missingness Model' estimator models the relationships between the covariates, $\X$ only. 

The z-, joint-, and no outcome models can in principle model all of the missingness structures present in simulation DAGs presented in Figure \ref{fig:sim dags}. This means that they are able to identify those MNAR mechanisms, but only when the parametric forms are correctly specified. As we will discuss in the simulation section, since these models specify full linear or linear up to a link function, relationships between variables can be misspecified with respect to the true data generating process when those are generated non-linearly. 

\subsection{Multiple imputation using chained equations (MICE)}
Multiple imputation using chained equations (MICE) \citep{van2011mice} is an example of a fully conditional multiple imputation procedure. The idea is to fit a model, implicitly or explicitly, for the missing data conditional on all observed cases and all other variables in such a way that many complete sets of data can be generated. In the case that there are several missing variables, the imputation is done sequentially, where latter variables are modelled conditional on both observed and imputed values of previous variables. For imputation of continuous variables we consider - Predictive Mean Matching (PMM) imputation, linear regression imputation, random forest imputation, and random indicator imputation. The first three MICE imputation approaches mentioned are created under the assumption of MAR data. Recent scholarship has shown that PMM, in particular, does have indentifiability guarantees even in non-parametric settings under MAR missingness when the stochastic imputation is used in conjunction with the fully conditional specification (FCS) approach as is default when using PMM-MICE for multiple imputation \citep{naf2024good}. Since we use a single imputation approach in this paper to compare more fairly with the machine learning approaches, the identifiability approaches may not hold. In addition, as pointed out in \cite{naf2024good}, identifiability does not imply consistency. The random indicator MICE procedure, on the other hand, is designed to handle some cases of MNAR data by estimating an offset between the observed and unobserved data \citep{jolani2012dual}. To our knowledge, formal identification results for the random indicator approach have not been proven.



\subsection{Machine Learning Imputation Methods}

\subsubsection{Partial Variational Auto-Encoder (PVAE)}
Variational Auto-Encoders \citep{kingma2013auto} (VAEs) are a class of generative models which augment the observed marginal data distribution with a continuous latent space denoted $z$. More recently, VAEs have been adapted to missing data problems, in particular a class of Partial VAEs \citep{ma2018eddi} (PVAEs) have been proposed for MAR missingness patterns. As is standard in the VAE literature, the PVAE minimizes the Evidence Lower Bound (ELBO) with respect to a tractable approximate:
\begin{align*}
\log \bp(\Xo) \leq \log \bp(\Xo)  - D_{KL}(q(z|\Xo)||\bp(z|\Xo)),
\end{align*}
where $\bp(\Xo)$ is the marginal distribution,$q(z|\Xo)$ is the tractable approximate posterior where both distributions are parameterized by a deep neural network architecture, and $D_{KL}(\cdot)$ is the K-L divergence. The key assumption of PVAEs is that $\Xo\indep\Xu|\Z$ which implies that the fully observed variables are sufficient for learning the latent generative variables $\Z$ since:
\begin{align*}
\bp(\Z|\Xo,\Xu) &= \frac{\bp(\Xo|\Xu\Z)\bp(\Xo|\Z)\bp(\Z)}{
\bp(\Xu|\Xo)\bp(\Xo)}\\
&= \frac{\bp(\Xo|\Z)\bp(\Z)}{
\bp(\Xo)} = \bp(\Z|\Xo).
\end{align*}
Learning the latent variables from the observed variables thus permits generation of the missing variables under the model assumptions, i.e. MAR missingness. Since the simulation DAGs in Figure \ref{fig:sim dags} feature self-censoring and are thus all examples of MNAR missingness mechanisms, PVAE, is theoretically unable to identify any of the simulation settings. We can think of this approach as being an MAR approximation to the MNAR problem with the added parametric flexibility afforded by the VAE structure.

\subsubsection{Not-missing-at-random Importance Weighted Auto-encoder (Not-MIWAE)}
Not-MIWAE \cite{Ipsen2020} is another VAE based generative imputation model which builds upon the missing data importance weighted auto-encoder (MIWAE) \citep{Mattei2019}, a generative model for MAR mechanism, and extends it to some cases of MNAR mechanism. Not-MIWAE uses the Importance weighted auto-encoder \citep{burda2015importance} loss function which is an extension to VAEs which aims to encourage more expressive generative networks by generating multiple realizations from the approximate posterior and averaging over them in the loss function with importance sampling techniques. It can be shown that the resulting lower bound is tighter than the VAE bound and approaches the marginal log-likelihood as the number of samples tends to infinity \citep{Ipsen2020}. The innovation with MIWAE compared to the PVAE approach is an explicit model for the missingness mechanism, similar to the joint Bayesian framework. In Not-MIWAE the missingness model may depend on partially observed data $\Xu$ and as a result can handle some classes of MNAR data. The key assumption is that the model is fully factorized which implies that $\bp(\Xo,\Xu|\Z) =\bp(\Xo|\Z)\bp(\Xu|\Z)$. Unlike the PVAE approach the missing data does not drop out of the augmented data and instead is integrated over along with the augmented latent space. The observed data likelihood can be written as:
\begin{align*}\nonumber
    \bp(\Xo,\M) = \iint & \bp(\M|\Xo,\Xu = \xu)\bp(\Xo|\Z = \z) \times \bp(\Xu|\Z = \z)d\xu d\z.
\end{align*}
In this approach the integral is done using importance sampling weights with the tractable approximate posterior. Since it is required to integrate over both the latent and missing data space, the reparametrization trick is used for both the approximate posterior and the missing data likelihood $\bp(\Xo|\Z)$ to perform back propagation.

Although Not-MIWAE is generated from a graphical model which corresponds to Figure \ref{fig:simdag1}, the VAE used in the Not-MIWAE architecture has not been shown to be identifiable \citep{Ma2021}. Therefore, although Not-MIWAE is designed to handle some cases of MNAR, it may not be identifiable and thus may not produce consistent estimates under any of the considered simulation settings.

\subsubsection{deep Generative Imputation Model for Missing Not At random (GINA)}
GINA \citep{Ma2021} is similar to Not-MIWAE in that it is a VAE approach for MNAR structures which explicitly models the missing mechanism. A difference with Not-MIWAE is the authors focus on the identifiability of the underlying VAE. Building off of identifiability conditions detail for VAEs and nonlinear Independence Component Analysis (ICA) \citep{Khemakhem2020}, the authors detail sufficient conditions for which MNAR missingness conditions can be identified with VAEs. These include the mild parametric restrictions to the latent component that they belong to conditional factorial exponential family. Further, they show that their proposed algorithm, GINA, meets these requirements. Under identifiability conditions, it can be shown that GINA can consistently estimate features of the observed data distribution for MNAR structures as complicated as those shown in Figure \ref{fig:simdag2}. 


\section{Simulations}\label{sc:sim}
 We develop a simulation framework to better understand the relationship between imputation methods and performance on the prediction error in the domain adaptation task and to compliment our theoretical understanding of the problem as described in Section \ref{sc:reduction}.

\subsection{Simulation Setup}
We compare performance of the different imputation methods outlined in Section \ref{sc:impute} with respect to recovering the missing values as well as in terms of prediction in the target domain. We generate the data using the family of structural equations shown in equations \ref{eqn::DGM}. To focus our attention on the imputation task and avoid introducing additional variables which depend more on the outcome model than the imputation procedures, we restrict our attention to the case where $g_x(\cdot)$, $h_x(\cdot)$, and $h_m(\cdot)$ are linear, that is the conditional outcome and the missingness indicator are linear functions of $\X$ and $\M$ on the link scale. We do allow $f_x(\cdot)$ to be non-linear and vary this non-linearity in the simulation. The structural equations are right enough to be compatible with each of the eight simulation DAGs represented in Figure \ref{fig:sim dags} (i.e with and without the direct paths from $\M$ to $\Y$). We consider the underlying data generating DAGs to be simulation parameters which are varied across simulation runs.

In addition to the generating DAGs, there are many parameters that we can vary within the context of the chosen structural equation and these parameters may interact in non-trivial and/or non-intuitive ways with the imputation procedures and the domain adaptation task. To handle this, we take a factorial design approach to design the simulation study, as shown in Table \ref{sim-table}. The factorial design allows us to infer both marginal simulation effects, i.e averaging over all simulation runs, and conditional effects, i.e conditioning on DAGs or other sets of simulation parameters in the sense of \cite{stokes2023simulation}.

\begin{table}[]
\centering
\begin{tabular}{@{}l|l@{}}
\toprule
Variable                             & Values                          \\ \midrule
Sample Size ($n$)                    & 750, 1000, 1500, 3000, 5000     \\
Target data Proportion               & 0.3, 0.5, 0.75                   \\
Missing Data Pct                     & 0.05, 0.1, 0.15, 0.2, 0.25, 0.3 \\
Dag                                  & 1, 2, 3, 4, 5, 6, 7, 8          \\
Number of layers in X (nonlinearity) & 1, 3, 5, 7                     \\ \bottomrule
\end{tabular} \vspace{1em}
\caption{Factorial design for the simulation study. In total there are 2880 possible combinations of the simulation variables and we run over 2300 of these combinations. For each of the 2300 combinations run, up to 200 simulations are performed with these settings. This results in approximately 120,000 different simulations where 18 different imputation methods are tested within each run. Other important variables such as the coefficients determining the strengths of the various causal relationships are generated randomly within each of the 120,000 simulations.}\label{sim-table}
\end{table}

 First, we vary the degree of non-linearity in the relationships between the covariates, represented by $f_x(\X_{1:(i-1)})$ in Equation \eqref{eqn::DGM}, from a linear function all the way to a neural network mean function with seven hidden layers and several nodes within each layer (ReLU activation functions and linear final output layer). For each DAG, we also vary the proportion of missingness in the partially observed variables (from 5\% to 30\%), and the number of observations in the source and target data(from $n = 500$ to $n = 3000$ where the target set contained between 30\% and 75\% of the number of samples in the source data). All of the simulation parameters are shown in Table \ref{sim-table}.

One of the important factors which is not set as an explicit simulation parameter is how strong the various causal relationships are in the respective data generating mechanisms. Since the strengths of the relationships are likely to interact in complicated ways and it would be difficult to parametrize the notion of strength in a meaningful and interpretable way without a very high dimensional parametrization, we opted to randomly generate the structural equation coefficients in each simulation run. In each simulation run the coefficients are drawn from a normal distribution with mean zero and standard deviation of 1.5. In this way our simulation results marginalize over a variety of causal strength settings.

\subsection{Model Fitting}
For clarity, we briefly describe the model fitting procedure used to produce predictions of the target outcome for each imputation method and simulation setting. First, we use the source and target data to generate imputed values independently in the two data sources. We use $(\Xo^{T},\Xu^T,\R^T)$ to generate $\hat{\Xu^T}$ and $(\Xo^{S},\Xu^S,\R^S)$ to generate $\hat{\Xu^S}$. Using the imputed data set $(\Xo^{T},\hat{\Xu^T},\R^T,\Xo^{S},\hat{\Xu^S},\R^S)$, the importance sampling weights are estimated using a classifier which takes in the imputed covariate set and missingness masks to predict which of the two data sources it belongs to, $\bp(Sample \in T|\Xo,\hat{\Xu},\R)$. The importance sampling weights are then constructed as $\bp(Sample \in T|\Xo,\hat{\Xu},\R)/\bp(Sample \in S|\Xo,\hat{\Xu},\R)$ which is proportional to $\bpt(\Xo,\Xu,\R)/\bps(\Xo,\Xu,\R)$. Using these importance weights we fit the conditional outcome model. In principle the conditional outcome model can be any model which admits a weighting. For simplicity, we fit the outcome model using a weighted Bayesian linear regression model in the simulation study. The target outcome predictions are then generated by using the imputed target data set $(\Xo^T, \hat{\Xu^T},\R^T)$ to generate conditional predictions from the reweighted outcome model, which we call $\hat{\Y}^T$.

In order to isolate the impact of the imputation procedures, we use the exact same weighting and conditional outcome estimation procedures for all imputation methods and simulation settings. In accordance with the theory developed in Section \ref{sc:reduction}, we evaluate the imputation procedures with respect to the domain adaptation task. That is, we primarily evaluate the RMSE of the target predictions and the true target values $(\sqrt{\frac{1}{n_t}\sum_{i=1}^{n_t}(\Y_i^T - \hat{\Y_i^T})^2}$. Since only the imputation methods vary in the simulations, the differences in performance of the domain adaptation procedure reflect differences in the imputation procedures only. As a robustness check, we additionally look at the RMSE for the imputation task alone for the covariates which were masked in the target.

\subsection{Overall Performance Marginalized Over Simulation Settings}

\begin{figure}[t]
\centering
\includegraphics[scale = 0.5]{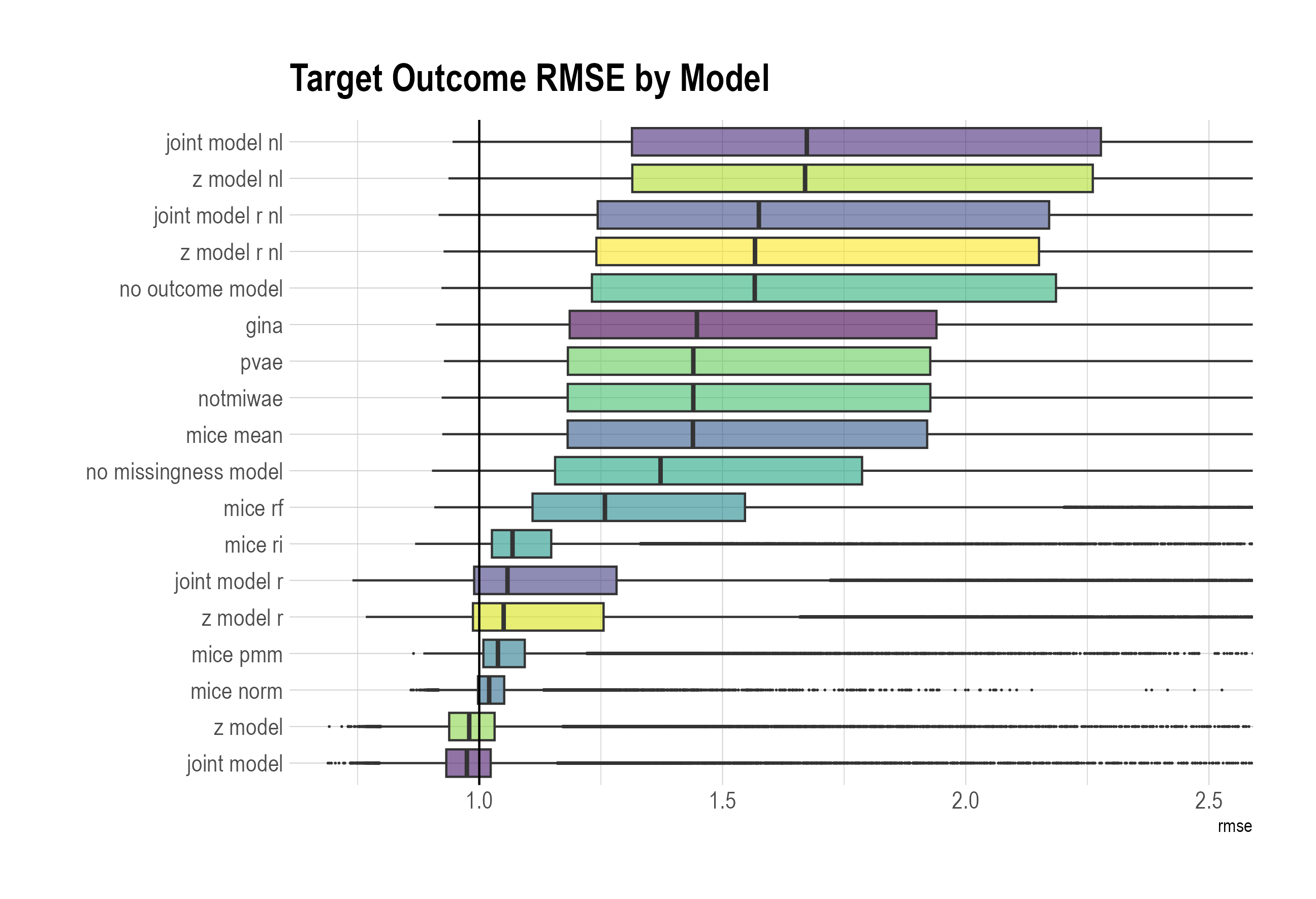}
\caption{Root mean squared errors (RMSE) in target domain adaptation for different imputation approaches marginalized over simulation settings shown in Table \ref{sim-table}. The imputation models fall into different categories: full Bayesian models with leakage of the target outcome (joint model and z model), full Bayesian models with no leakage of the target outcome  (joint model nl and z model nl), partial Bayesian models with leakage (z model r and joint model r), partial Bayesian models with no leakage (z model r nl and join model r nl), different implementations of mice as a single imputation (mice pmm, mice norm, mice rf and mice ri), and machine learning models (gina, pvae and notmiwae). As a benchmark, we also include mean imputation (mice mean).}  \label{fig:marg-plot}
\end{figure}

For each imputation model, we first present RMSE marginalized over different simulation parameters in Figure \ref{fig:marg-plot}. 
The imputation models fall into different categories as presented in Section \ref{sc:impute}, including full Bayesian models with and without leakage of the target outcome, partial Bayesian models, mice, and machine learning models. As a baseline approach, we also include results from mean imputation where we replace missing covariates with their corresponding mean.



In Figure \ref{fig:marg-plot} we can see that the RMSE are well organized by the imputation model categories. In particular, the Full Bayesian models which leak the target outcome information  (z model and joint model) perform best. When the joint models are reasonably well specified, there should be information to gain with respect to the imputation task from the outcome model as well. Notice, however, that when we do not use the target outcome information these models (joint model nl and z model nl) perform worse than all other model categories. Despite the fact that the joint Bayesian models with no leakage (joint model nl and z model nl) use additional information for the source imputation, they perform similarly to the no outcome models.

Similarly, the partial Bayesian models without leakage that do not explicitly model outcomes (joint model r nl and z model r nl) do not perform well and are comparable to the joint models without leakage. While the partial Bayesian models with leakage (z model r and joint model r) have smaller RMSE than the partial Bayesian models without leakage, they have higher RMSE than the full Bayesian models with leakage. This may be because of the notion of congeniality in imputation, where an imputation model is said to be congenial if there exists an underlying Bayesian model which links the imputation model and the model for the analysis \citep{d2024behind}. In this case, by excluding the information from the outcome model in the imputation, the imputation model explicitly does not condition on the outcome information and thus is not congenial. This is important when using Bayesian models for imputation or more generally when using stochastic imputation procedures \citep{d2024behind}. One way to mimic this broader underlying Bayesian model is to include the outcome directly in the imputation model, but as discussed this may be inappropriate in some Domain Adaptation settings, including the one we focus on in this manuscript, where it is assumed that we do not have access to the outcome information from target domain.

 The various mice implementations, the predictive mean matching (PMM) and linear regression in particular, perform well out of the box, with comparable RMSE to the joint Bayesian models with leakage. In these simulations, we use a single best imputation from the mice models as our imputation model rather than generating multiple data sets to perform multiple imputation as is common with the mice framework. This was to more easily compare with some of the other frameworks, in particular the machine learning methods, which would have required adaptations to generate multiple data sets. Thus, the mice imputation in our context are examples of deterministic imputation. It has been shown that in cases of deterministic imputation, excluding the information from the outcome is preferred since its inclusion can lead to biased imputation estimates \citep{d2024behind}. This may explain the performance gap between the MICE models and the Partially Bayesian models which share similar features otherwise.

\subsection{Performance for Specific Simulation Settings}
\begin{figure}[h]
\includegraphics[scale = 0.4]{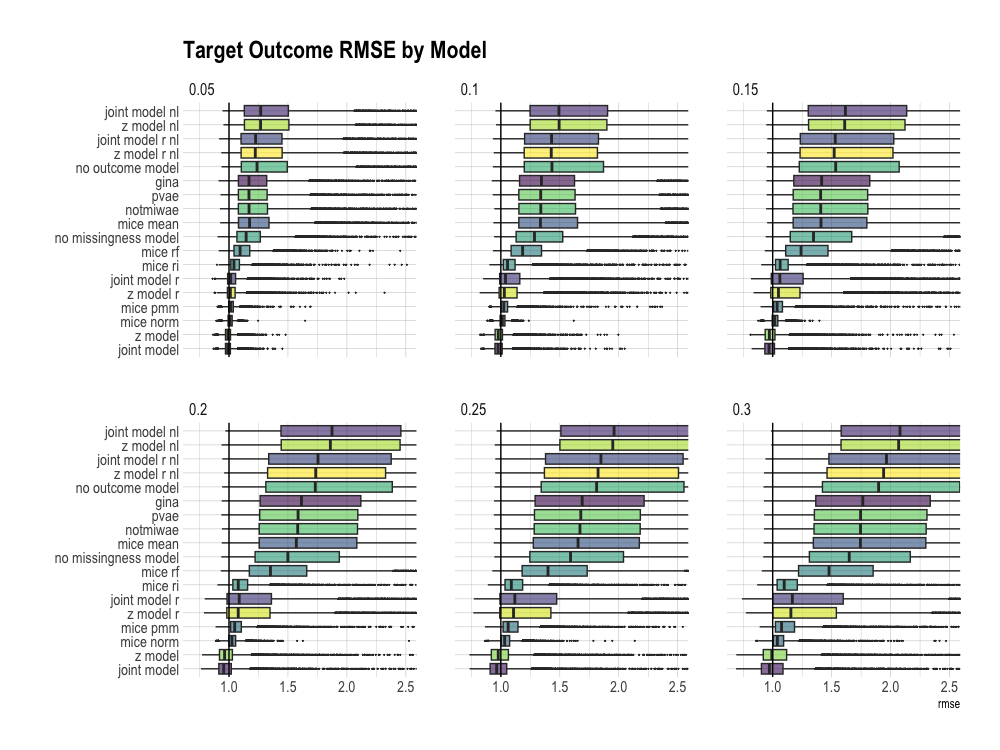}
\caption{Performance of various imputation models with respect to root mean squared error (RMSE) for target domain adaptation stratified by missingness proportions (0.05, 0.1, 0.15, 0,2,, 0.25, and 0.3) in the simulation study. The imputation models fall into different categories: full Bayesian models with leakage of the target outcome (joint model and z model), full Bayesian models with no leakage of the target outcome  (joint model nl and z model nl), partial Bayesian models with leakage (z model r and joint model r), partial Bayesian models with no leakage (z model r nl and join model r nl), different implementations of mice as a single imputation (mice pmm, mice norm, mice rf and mice ri), and machine learning models (gina, pvae and notmiwae). As a benchmark, we also include mean imputation (mice mean).}\label{fig:marg-miss}
\end{figure}

We also explore the performance of the models by different simulation setting. First, we explore the performance over simulation settings with different missingness probabilities, ranging from 5\% to 30\%. As seen in Figure \ref{fig:marg-miss}, while the broad trends in RMSE in the imputation models are largely unaffected within each missingness probability, the overall performance decreases with increased RMSE as the missingness probability increases from 5\% to 30\%.

Second, we explore the influence of the data generating DAGs (Figure \ref{fig:sim dags}) on the simulation results. 
As shown in Figures \ref{fig:marg-14} and \ref{fig:marg-58} in the Supplementary Materials, simulation DAGs 2, 3, 6 and 7 result in on average higher imputation errors and higher variance compared to other DAGs. Further, some of the ordering of the imputation methods with respect to their performances, varies by the DAG. For example, Gina performs slightly better relative to the other machine learning methods in DAGs 5-8 where there is a causal relationship between the missingness indicators and the outcome, this is particularly true in DAGs 6 and 7. However, such effects tend to be small relative to the overall trends with respect to the imputation model categories.

Finally, we assess the role of the non-linearity in the relationship between the covariates in imputation performance. In Figure \ref{fig:marg-cat-nl} shown in the Supplementary Material, we see that the machine learning methods perform relatively better compared to the partially Bayesian and mean imputation methods as the relationship between the covariates becomes more nonlinear, where 1 means that this relationship is generated by a single-layer neural network with 1 node and 7 means that the relationship is generated by a 7 layer neural network with 7 nodes each. This confirms intuition that imputation models which posit more non-linear functional forms perform relatively better as the data generating mechanism becomes more non-linear. However, the effect in these simulations is not so large as to give the machine learning imputation a significant advantage over the better performing Bayesian and MICE models. It is possible that a higher covariate dimension and hyper-parameter optimization would be necessary to see such results, which we intend to explore in future work.

\subsection{Relationship Between Imputation Error and Domain Adaptation Problem}

In Section \ref{sc:reduction} we show that the domain adaptation problem can be reduced to an imputation error. To empirically verify if this theorem holds, we first assess whether imputation error in the target (imputation RMSE in target) is predictive of domain adaptation prediction error (target outcome RMSE) in the simulations across difference simulation settings and imputation methods. Indirectly, the domain adaptation error is also a function of the source imputation error. In the case of MNAR missingness, the estimation of the conditional expectation will be a function of the source imputation error which will impact the ability for the source prediction to transfer to the target set. Theorem \ref{thm:1} assumes a conditional expectation estimator which converges asymptotically to the true conditional expectation, but in practice in finite samples it is an open question as to whether the source imputation error or target imputation error will dominate in determining the target domain adaptation prediction error. Using our simulations, we further build a regression model to determine the relative importance of the two imputation errors. To do so we use a Generalized Additive Model (GAM) with smoothed terms through target imputation error, source imputation error, and their interaction term.

\begin{figure}[h]
    \centering
\includegraphics[scale = 0.3]{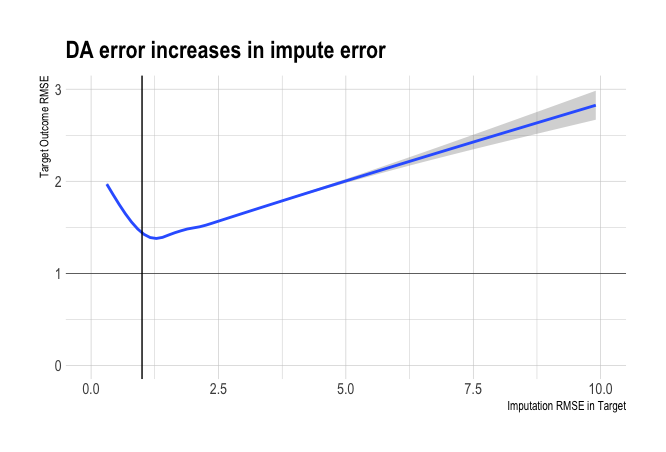}
    \caption{Relationship between imputation error in the target and domain adaptation prediction task. This figure uses all simulation results and all imputation methods.}\label{fig:impute error}
\end{figure}

In Figure \ref{fig:impute error}, we can first see that empirically the domain adaptation error increases monotonically (on average) in the target imputation error above the empirical risk minimum imputation error which was set to 1  for the simulation study. In some cases, the imputation error was below 1 and occasionally substantially so. We can see that imputing better than the empirical risk minimum does not in general lead to better domain adaptation prediction results. This is likely because such imputation performances were achieved by chance and not indicative of estimating something closer to a random variable which converges in probability to the missing data. Thus the simulation results depicted in Figure \ref{fig:impute error} concur with the theoretical results in Theorem \ref{thm:1} which shows the domain adaptation error as a function of the target imputation error.

\begin{figure*}
\centering
\begin{subfigure}{.4\textwidth}
\scalebox{.25}{
 \includegraphics{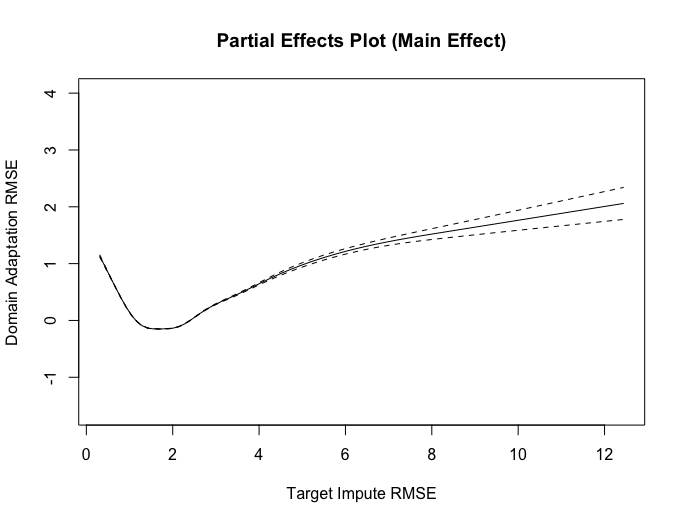}
 }
 \caption{}\label{fig:par-target}
 \end{subfigure}
\begin{subfigure}{.4\textwidth}
\scalebox{.25}{
 \includegraphics{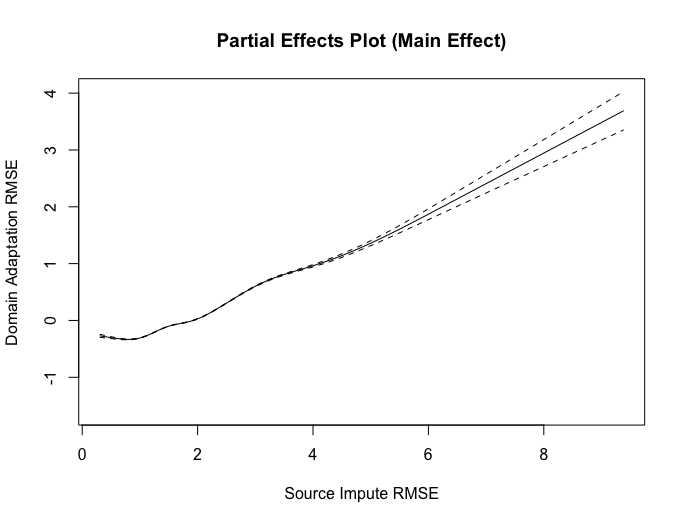}
 }
\caption{}\label{fig:par-source}
 \end{subfigure}
 \begin{subfigure}{.4\textwidth}
\scalebox{.25}{
 \includegraphics{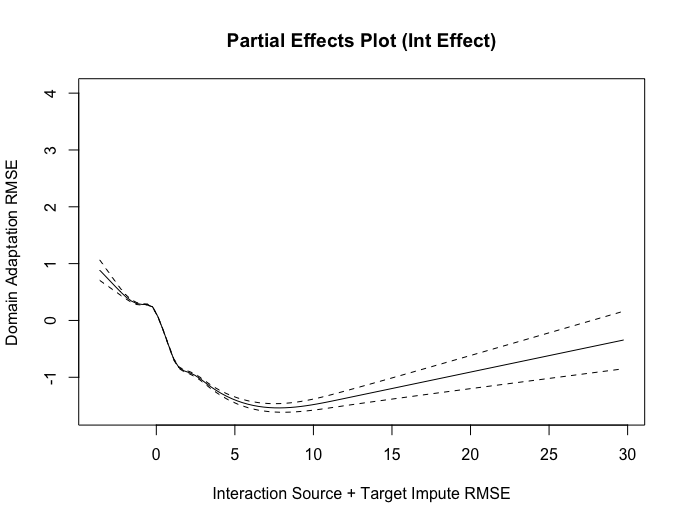}
 }
\caption{}\label{fig:par-int}
 \end{subfigure}
 \caption{Influence of imputation errors (target, source and their interactions) on domain adaptation prediction task. Along the y-axis is domain adaptation root mean squared error (RMSE) and along the x-axis is target imputation RMSE (a), source imputation RMSE (b), and their sum (c). The regression curves in each sub-figure are based on generalized additive model with smoothed terms through target imputation error, source imputation error, and their interaction term. This figure uses all simulation results and all imputation methods.}\label{fig:partial}
\end{figure*}

In Figure \ref{fig:partial} we show the partial effects plots of the target and source RMSE, respectively, as well as the interaction effect between the two. The main effect of the source imputation error is the most predictive with respect to the F-statistic. This indicates that the role of the source imputation on the quality of estimation of the conditional expectation may be as important or more important than target imputation error in practice with finite samples. These patterns hold when controlling for the missingness probabilities and sample sizes in the simulation.

\section{Application to eICU EHR data}\label{sc:application}

We apply the general methodology from Section \ref{sc:reduction} to a real world EHR data from the Philips eICU program, \cite{pollard2018eicu}, a publicly available collaborative research database (CRD) containing records from intensive care units (ICUs) throughout the United States. Our primary objective is to predict ICU mortality risk using the extracted features, which encompass demographic information, physiological data, and laboratory results. To transform this into a domain adaptation problem we split the data into source and target domains based on geographical region, as it has been reported that there is a distributional shift as well as a generalization gap between the regions \citep{singh2022generalizability}.  We consider the Southern region as the source and the Northeast region as the target. 

In total, we extracted 57 features from vital signs, laboratory measurements, demographic information (age, gender, and race/ethnicity), and an indicator for elective surgery admission.  We removed variables with a high proportion of missing values (more than 70\%). The vital signs and laboratory measurements in the data set were recorded repeatedly. For these variables, we extracted data within fixed time windows of 24 hours and 48 hours. Within the 24-hour window, we derived key features such as the first, last, minimum, and maximum values for heart rate, systolic/diastolic/mean blood pressure, respiratory rate, temperature, and oxygen saturation (SaO$_2$) from vital measurements. In the 48-hour window, we similarly calculated first, last, minimum, and maximum values for base excess, calcium, partial pressure of oxygen in arterial blood (PaO$_2$), partial pressure of carbon dioxide in arterial blood (PaCO$_2$), glucose, pH, the ratio of PaO$_2$ to fraction of oxygen inspired (PaO$_2$/FiO$_2$ ratio), albumin, blood urea nitrogen (BUN), immature band forms (bands), bicarbonate, bilirubin, creatinine, hematocrit, hemoglobin, lactate, platelet count, potassium, international normalized ratio (INR), sodium, and white blood cell count (WBC) from laboratory measurements. The total output variable was obtained by summing repeated measures within the 24-hour time window.

The eICU-CRD dataset encompasses 200,859 distinct stays in ICU units, involving 139,367 unique patients across 208 hospitals, spanning the years 2014 to 2015. We applied the APACHE IVa criteria as exclusion criteria \citep{zimmerman2006apache} and specifically considered only the initial ICU admission for each patient to prevent double-counting for the same individual \citep{singh2022generalizability}. Additionally, we excluded stays with unknown regions and patients under the age of 16. Following the data exclusions, we retained a total of 111,669 ICU stays from unique patients across 171 hospitals with overall ICU mortality rate of 5.88\%. Our analytical data includes 8,158 patients from the Northeast and 39,433 patients from the Southern regions, with a total of 57 covariates. 

 We first summarize distribution of covariates and their missingness across the two regions. To estimate the importance sampling weights we trained a Xgboost model predicting whether the data came from the South or the Northeast region as a function of all the covariates post-imputation and the missingness patterns. We can use the estimated weights or propensities to better understand the similarity or lack there of in terms of covariate and missingness patterns between the source and target in terms of the propensity overlap. We show an example of the estimated propensity to be in the South or Northeast under simple mean imputation. 
 
To compare various imputation models for MNAR domain adaptation, we broadly follow the same steps as described in the simulation study in Section \ref{sc:sim}. That is, first we fit separate imputation models on the target and source data. Given the computational complexity in implementing the joint Bayesian models with high dimensional data, we do not use the Bayesian models for imputation in this application and only focus on comparing different mice and machine learning imputation approaches along with mean imputation. Second, we use the imputed data and the missingness labels to estimate the importance sampling weights with an Xgboost model. Finally, we fit the outcome model on the source data with the estimated importance sampling weights using a random forest model. We repeat the above process five different times with different seeds to ensure that the results are not overly explained by random seed luck. Finally, we compare the Brier score or mean squared error (MSE) and area under the receiver operating curve (AUROC) for prediction in the target data after imputation and adjusting for covariate and missingness shifts using importance weights across the different methods. Lower RMSE and higher AUROC indicate better performance with respect to target domain adaptation.

\subsection{Result} 
The analytical sample for the case study includes 39,433 patients from the South and 8,158 from the Northeast regions. As noted in \citet{singh2022generalizability}, there are notable differences in the distribution of demographic and clinical information across the two regions. For example, while 91\% of the patients in Northeast self-identified as White, only 68\% identify as White in South. Similarly, the percent of Black patients in the South region is 19\% compared to 3\% in Northeast, and slightly higher proportion of patients in Northeast had undergone elective surgery compared to those in the South (8.5\% vs 6.8\%). Furthermore, the distribution of missingness proportion of the covariates is also different in the two regions (Figure \ref{fig:miss-dist}). On average, there is a higher proportion of missingness in the South (17.9\%) compared to the Northeast (13.4\%). Further, the number of covariates with zero or close to zero missingness in the Northeast is much higher than that in the South (35\% vs 5\%). 


\begin{figure}[tb]
    \centering
    \includegraphics[scale = 0.5]{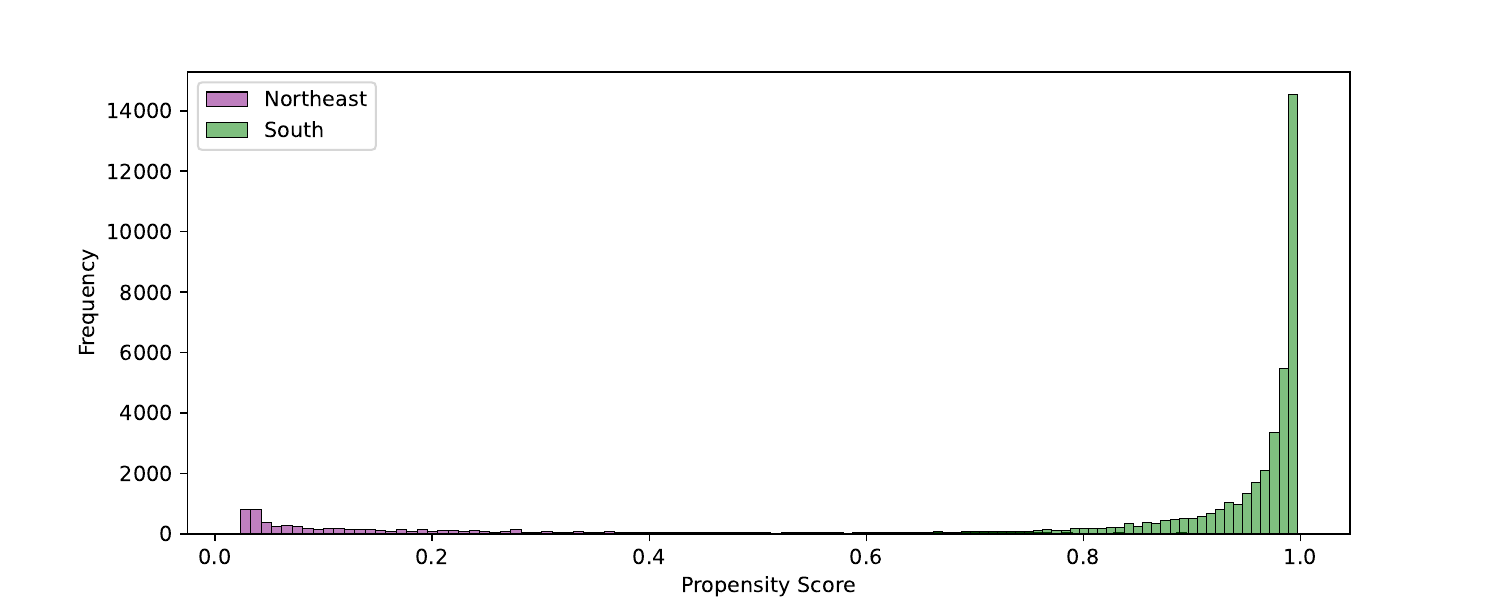}   
    \caption{Distribution of the estimated importance weights representing the probability that an observation belongs to the South Region demonstrating minimal overlap in covariates between the two regions. For this example, mean imputation was used to impute missing variables in both Northeast and South regions.}
    \label{fig:prop}
\end{figure}

 The histogram in Figure \ref{fig:prop} shows the distribution of the importance sampling weight or the propensity score (conditional predicted probabilities) of a patient coming from the South region (source) for mean imputation, as an example. As demonstrated by the separation of the two colors in the histogram (red for the Northeast and blue for the South), this classifier is able to well separate the two data sets with minimal overlap. In general, observations from the South have propensity scores near 1 (mean = 0.95) and observations from the Northeast region have propensity scores closer to 0 (mean = 0.19).

\begin{figure}[t]
    \centering
    \begin{subfigure}{0.45\textwidth}
        \centering
        \includegraphics[width=\textwidth]{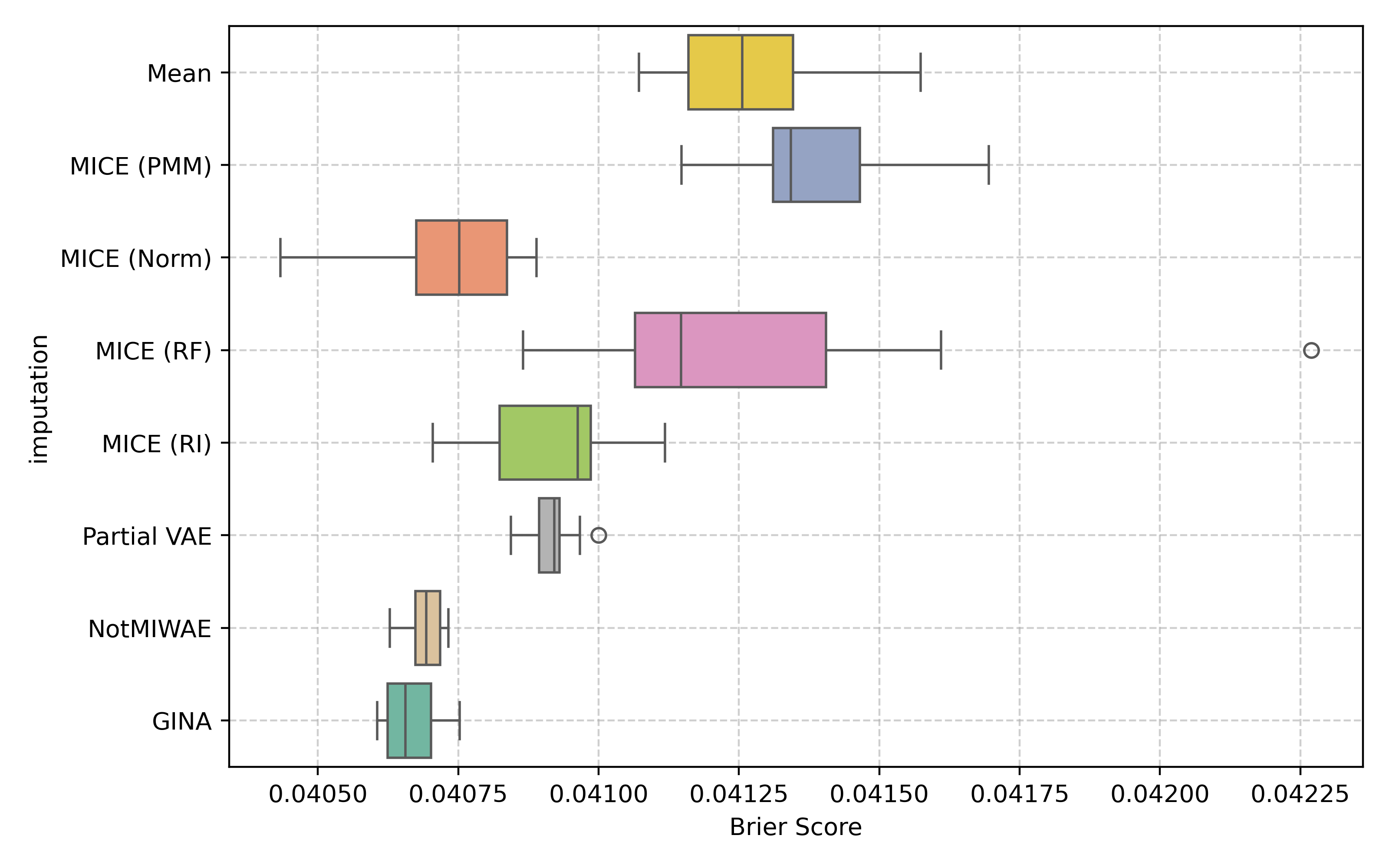}
        \caption{Brier Score (Mean Squared Error)}
        \label{fig:sub1}
    \end{subfigure}
    \hfill
    \begin{subfigure}{0.45\textwidth}
        \centering
        \includegraphics[width=\textwidth]{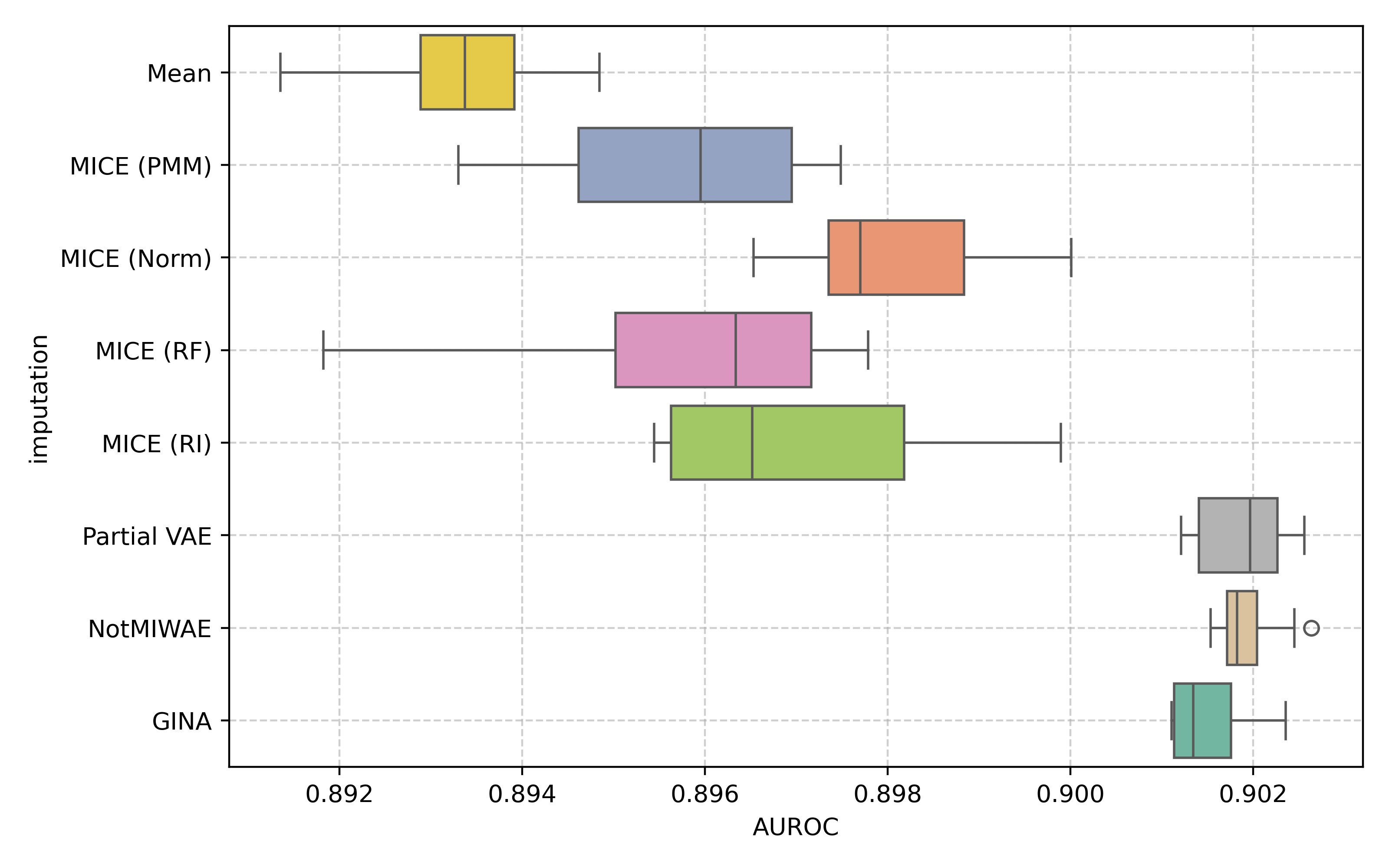}
        \caption{AUROC}
        \label{fig:sub2}
    \end{subfigure}
    \caption{Target domain adaptation performance with respect to (a) mean squared error and (b) AUROC for various imputation methods in EHR data application to predict ICU mortality in the South region using source data from Northeast.}
    \label{fig:result}
\end{figure}

The performance of the domain adaptation task, where we predict outcomes in the target dataset (South) after imputing missing covariates using various methods (Section \ref{sc:impute}), is demonstrated in Figures \ref{fig:result}. As seen in these figures, the machine learning-based imputation methods (Partial VAE, NotMIWAE, and GINA) generally outperform traditional MICE-based approaches in terms of both Brier Score and AUROC. Among multiple imputation methods, mice with linear regression imputation performs generally better. It also has comparable performance to partial VAE in terms of Brier score, though has lower AUROC. Finally, mean imputation in general has the worse performance.

Our results suggest that machine learning approaches may be better suited to handling high-dimensional imputation problems and capturing complex dependencies among variables. The relationships between the variables may be more complicated than in the simulation results favoring the more flexible machine learning methods. Additionally, the observed performance differences may be attributed to underlying distributional or concept shifts, i.e. shifts in the conditional distribution of the outcome given the predictors. Therefore, in addition to the covariate and missingness shifts, there may also be concept shift between hopspital regions as previously discussed by \citet{singh2022generalizability}. In such cases, the interaction between target outcome model misspecification and imputation methods is non-trivial, further emphasizing the advantages of more flexible machine learning approaches in adapting to real-world data complexities.

\section{Practical considerations and connection with causal inference literature}

One of the key assumptions throughout this work is that the importance weights are well defined in the sense that the ratio $\bpt(\Xo,\Xu,\R)/\bps(\Xo,\Xu,\R)$ exists for all sets $\mathcal{A}$ such that $\bps(\mathcal{A};\Xo,\Xu,\R)>0$. Asymptotically, this assumption may be relatively weak, but drawing analogies to the positivity assumption in the causal inference literature \citep{cole2008constructing,xiao2013comparison}, there may be practical violations in finite samples which can cause problems. This can be illustrated using the much simpler covariate shift example depicted in Figure \ref{fig:weight-simulation}. In this context, the importance sampling weights exist asymptotically since both $x_s \sim N(0,1)$
 and $x_t \sim N(1,1)$ have positive measures over all intervals on the real line. However, we can see that in the example with finite samples there are parts of the covariate space in the figure as we move away from 0.5 in either direction where there is concentration of samples from either source or target samples but not both. The result is that although the importance sampling weights tilt the source approximation towards the desired target approximation, this correction is not enough to eliminate the source bias. This is a problem that would go away asymptotically with infinite samples, but not with repeated sampling of finite samples without additional assumptions such as assuming a parametrically correct functional form for the conditional outcome.

In practical machine learning problems, in particular with the added concern of missing data, practical positivity violations of the importance sampling weights is likely to be much greater than the one-dimensional toy covariate shift problem shown in Figure \ref{fig:weight-simulation}. This is demonstrated in our real data example in Figure \ref{fig:prop} where we display the components of the importance sampling weights in the form of the probability that the covariates, imputed data, and missingness patterns belong to the source or target. We can see that most of the data is strongly classified as either coming from the source or the target with few rows plausibly belonging to either. In practice, this can lead to an even more extreme form of under-correcting the source approximation due to the lack of practical overlap in the very high-dimensional observed covariate plus missingness sample space.

An analogous type of under-smoothing in finite sampling has been recognized recently in the influence function literature, particularly in the challenging context of estimating conditional average treatment effect (CATE) functions \citep{kennedy2023towards}. In this context, so-called doubly robust estimators have been shown to perform significantly better than simple plug-in estimators which are susceptible to noise over parts of the sample space with insufficient treated or un-treated samples \citep{kennedy2023towards}. Many such doubly robust estimators can be viewed as a plug-in estimator plus a bias correction term which disappears asymptotically. In other words, they use the structure of the asymptotical problem to inform a finite sample correction which may reduce those finite sample biases as well as reduce the variance of the resulting estimators. 

Additionally, recent work in this broader literature has made progress understanding influence functions for risk minimizing functions themselves \citep{van2024combining}. By viewing the minimization problem as a functional itself, the principles of doubly robust estimators can be directly applied to empirical risk estimate problems like the domain adaptation minimization problem shown in equation \eqref{eq:minimizer}. This literature primarily focuses on estimating the CATE function in classic causal inference problems and we leave it to future work to determine the extent to which this framework can be applied to the domain adaptation framework.

Another potential limitation to the framework for practical applications is the strictness of the conditional outcome assumption, i.e the conditional outcome distribution is equivalent in the source and target distributions. In our simulations and EHR data application, unlike the canonical domain adaptation setup, we have access to the target data outcome labels which allowed us to run additional tests not feasible in stricter setups. In particular, we found that by training outcome models separately on the training and source data, we were able to significantly reduce the target error confirming results from \cite{singh2022generalizability} which suggested that the outcome-covariates relationships were different across different regions. The authors hypothesize  that the underlying distributions of the source and target distribution may be stronger of an assumption than strictly necessary to improve the source approximation towards the desired target approximation. In future work, we will explore the extent to which weaker assumptions directly on the conditional outcome expectation, and smoothness thereof, may be combined with assumptions of overlap to achieve a weaker form of domain adaptation which approximates well over the covariate space of the target domain. We may be able to borrow concepts again from the causal inference literature and define a new target estimator over the regions in the covariate space with high overlap in the spirit of the average treatment effect in the overlap \citep{li2019propensity}. At a minimum, such a target estimator would only depend on the conditional outcome distributions over the overlap space, but there may be ways to further weaken these results by focusing on particular moments of the conditional outcome distribution over the overlap space such as the conditional mean.



\section{Conclusion}
In this paper, we show that the problem of domain adaptation under MNAR missingness shift can be reduced to imputing the missing data in both the source and the target separately. By combining modern MNAR missingness imputation techniques, whereby we impute both source and target data, and classic tools from the domain adaptation for covariate shift literature, such as re-weighting with importance sampling weights, we extend missingness shift to a subset of MNAR structures. We further provide a detailed review of various traditional statistics and machine learning methods capable of handling MNAR missingness under a set of identifying assumptions.

Through extensive simulation study and a real world data example with EHR data from hospitals at two different geographic regions in the US, we assess performance of the target domain adaption under MNAR shift for these different imputation models. While the imputations using joint Bayesian models and mice demonstrate superior performance in the simulation study, modern machine learning methods perform better in the real data application. In low-dimensional simulation settings, parametric joint Bayesian models which directly model the dependence structures in the missing DAG tend to perform better than more complex machine learning models. Our simulations show that by additionally modeling the outcome distribution simultaneously with missingness there is gain in imputation performance. This perhaps motivates a natural extension of the more flexible MNAR imputation methods like GINA which attempts to simultaneously model the joint distribution over the covariates, missingness and outcome. 
In the EHR application, possibly with more complicated assumptions beyond our considerations, machine learning approaches perform better than restrictive parametric methods. However, it is important to note that none of the methods out-perform mean imputation by a large margin. This highlights a methodology gap with respect to real-world domain adaptation under MNAR missingness shift. 
\clearpage

\section*{Acknowledgements}\label{acknowledgements}
The authors gratefully acknowledge support from NIH grant R01HL155149.

\section*{Conflict of interest}



\bibliographystyle{unsrtnat}
\bibliography{bib, supplement_bib}

\begin{thebibliography}{34}
\providecommand{\natexlab}[1]{#1}
\providecommand{\url}[1]{\texttt{#1}}
\expandafter\ifx\csname urlstyle\endcsname\relax
  \providecommand{\doi}[1]{doi: #1}\else
  \providecommand{\doi}{doi: \begingroup \urlstyle{rm}\Url}\fi

\bibitem[Ben-David et~al.(2010)Ben-David, Blitzer, Crammer, Kulesza, Pereira,
  and Vaughan]{ben2010theory}
Shai Ben-David, John Blitzer, Koby Crammer, Alex Kulesza, Fernando Pereira, and
  Jennifer~Wortman Vaughan.
\newblock A theory of learning from different domains.
\newblock \emph{Machine learning}, 79:\penalty0 151--175, 2010.

\bibitem[Ben-David et~al.(2006)Ben-David, Blitzer, Crammer, and
  Pereira]{ben2006analysis}
Shai Ben-David, John Blitzer, Koby Crammer, and Fernando Pereira.
\newblock Analysis of representations for domain adaptation.
\newblock \emph{Advances in neural information processing systems}, 19, 2006.

\bibitem[Gretton et~al.(2009)Gretton, Smola, Huang, Schmittfull, Borgwardt,
  Sch{\"o}lkopf, et~al.]{gretton2009covariate}
Arthur Gretton, Alex Smola, Jiayuan Huang, Marcel Schmittfull, Karsten
  Borgwardt, Bernhard Sch{\"o}lkopf, et~al.
\newblock Covariate shift by kernel mean matching.
\newblock \emph{Dataset shift in machine learning}, 3\penalty0 (4):\penalty0 5,
  2009.

\bibitem[Zhang et~al.(2013)Zhang, Sch{\"o}lkopf, Muandet, and
  Wang]{zhang2013domain}
Kun Zhang, Bernhard Sch{\"o}lkopf, Krikamol Muandet, and Zhikun Wang.
\newblock Domain adaptation under target and conditional shift.
\newblock In \emph{International conference on machine learning}, pages
  819--827. PMLR, 2013.

\bibitem[Nair et~al.(2019)Nair, Satpathy, Christopher,
  et~al.]{nair2019covariate}
Nimisha~G Nair, Pallavi Satpathy, Jabez Christopher, et~al.
\newblock Covariate shift: A review and analysis on classifiers.
\newblock In \emph{2019 Global Conference for Advancement in Technology
  (GCAT)}, pages 1--6. IEEE, 2019.

\bibitem[Gama et~al.(2014)Gama, {\v{Z}}liobait{\.e}, Bifet, Pechenizkiy, and
  Bouchachia]{gama2014survey}
Jo{\~a}o Gama, Indr{\.e} {\v{Z}}liobait{\.e}, Albert Bifet, Mykola Pechenizkiy,
  and Abdelhamid Bouchachia.
\newblock A survey on concept drift adaptation.
\newblock \emph{ACM computing surveys (CSUR)}, 46\penalty0 (4):\penalty0 1--37,
  2014.

\bibitem[Garg et~al.(2020)Garg, Wu, Balakrishnan, and Lipton]{garg2020unified}
Saurabh Garg, Yifan Wu, Sivaraman Balakrishnan, and Zachary Lipton.
\newblock A unified view of label shift estimation.
\newblock \emph{Advances in Neural Information Processing Systems},
  33:\penalty0 3290--3300, 2020.

\bibitem[Rubin(1976)]{rubin1976inference}
Donald~B Rubin.
\newblock Inference and missing data.
\newblock \emph{Biometrika}, 63\penalty0 (3):\penalty0 581--592, 1976.

\bibitem[Zhou et~al.(2023)Zhou, Balakrishnan, and Lipton]{zhou2023domain}
Helen Zhou, Sivaraman Balakrishnan, and Zachary Lipton.
\newblock Domain adaptation under missingness shift.
\newblock In \emph{International Conference on Artificial Intelligence and
  Statistics}, pages 9577--9606. PMLR, 2023.

\bibitem[Pollard et~al.(2018)Pollard, Johnson, Raffa, Celi, Mark, and
  Badawi]{pollard2018eicu}
Tom~J Pollard, Alistair~EW Johnson, Jesse~D Raffa, Leo~A Celi, Roger~G Mark,
  and Omar Badawi.
\newblock The eicu collaborative research database, a freely available
  multi-center database for critical care research.
\newblock \emph{Scientific data}, 5\penalty0 (1):\penalty0 1--13, 2018.

\bibitem[Little and Rubin(2019)]{Little2019}
Roderick J~A Little and Donald~B Rubin.
\newblock \emph{Statistical analysis with missing data}, volume 793.
\newblock John Wiley \& Sons, 2019.

\bibitem[Mohan et~al.(2013)Mohan, Pearl, and Tian]{Mohan2013}
Karthika Mohan, Judea Pearl, and Jin Tian.
\newblock Graphical models for inference with missing data.
\newblock \emph{Advances in neural information processing systems}, 26, 2013.

\bibitem[Bhattacharya et~al.(2020)Bhattacharya, Nabi, Shpitser, and
  Robins]{Bhattacharya2020}
Rohit Bhattacharya, Razieh Nabi, Ilya Shpitser, and James~M Robins.
\newblock Identification in missing data models represented by directed acyclic
  graphs.
\newblock In \emph{Uncertainty in Artificial Intelligence}, pages 1149--1158.
  PMLR, 2020.

\bibitem[Nabi et~al.(2020)Nabi, Bhattacharya, and Shpitser]{Nabi2020}
Razieh Nabi, Rohit Bhattacharya, and Ilya Shpitser.
\newblock Full law identification in graphical models of missing data:
  Completeness results.
\newblock pages 7153--7163, 2020.

\bibitem[Ipsen et~al.(2021)Ipsen, Mattei, and Frellsen]{Ipsen2020}
Niels~Bruun Ipsen, Pierre-Alexandre Mattei, and Jes Frellsen.
\newblock not-miwae: Deep generative modelling with missing not at random data.
\newblock In \emph{International Conference on Learning Representations}, 2021.

\bibitem[Ma and Zhang(2021)]{Ma2021}
Chao Ma and Cheng Zhang.
\newblock Identifiable generative models for missing not at random data
  imputation.
\newblock \emph{Advances in Neural Information Processing Systems},
  34:\penalty0 27645--27658, 2021.

\bibitem[Du et~al.(2022)Du, Enders, Keller, Bradbury, and Karney]{Du2022}
Han Du, Craig Enders, Brian~Tinnell Keller, Thomas~N Bradbury, and Benjamin~R
  Karney.
\newblock A bayesian latent variable selection model for nonignorable
  missingness.
\newblock \emph{Multivariate Behavioral Research}, 57:\penalty0 478--512, 2022.

\bibitem[Van~Buuren and Groothuis-Oudshoorn(2011)]{van2011mice}
Stef Van~Buuren and Karin Groothuis-Oudshoorn.
\newblock mice: Multivariate imputation by chained equations in r.
\newblock \emph{Journal of statistical software}, 45:\penalty0 1--67, 2011.

\bibitem[D’Agostino~McGowan et~al.(2024)D’Agostino~McGowan, Lotspeich, and
  Hepler]{d2024behind}
Lucy D’Agostino~McGowan, Sarah~C Lotspeich, and Staci~A Hepler.
\newblock The “why” behind including “y” in your imputation model<?
  show [aq id= gq6]?>.
\newblock \emph{Statistical Methods in Medical Research}, page
  09622802241244608, 2024.

\bibitem[N{\"a}f and Josse(2024)]{naf2024good}
Jeffrey N{\"a}f and Julie Josse.
\newblock What is a good imputation under mar missingness?
\newblock \emph{arXiv preprint arXiv:2403.19196}, 2024.

\bibitem[Jolani et~al.(2012)]{jolani2012dual}
Shahab Jolani et~al.
\newblock \emph{Dual imputation strategies for analyzing incomplete data}.
\newblock Utrecht University, 2012.

\bibitem[Kingma and Welling(2013)]{kingma2013auto}
Diederik~P Kingma and Max Welling.
\newblock Auto-encoding variational bayes.
\newblock \emph{arXiv preprint arXiv:1312.6114}, 2013.

\bibitem[Ma et~al.(2019)Ma, Tschiatschek, Palla, Hernandez-Lobato, Nowozin, and
  Zhang]{ma2018eddi}
Chao Ma, Sebastian Tschiatschek, Konstantina Palla, Jose~Miguel
  Hernandez-Lobato, Sebastian Nowozin, and Cheng Zhang.
\newblock Eddi: Efficient dynamic discovery of high-value information with
  partial vae.
\newblock In \emph{International Conference on Machine Learning}, pages
  4234--4243. PMLR, 2019.

\bibitem[Mattei and Frellsen(2019)]{Mattei2019}
Pierre-Alexandre Mattei and Jes Frellsen.
\newblock Miwae: Deep generative modelling and imputation of incomplete data
  sets.
\newblock pages 4413--4423, 2019.

\bibitem[Burda et~al.(2015)Burda, Grosse, and
  Salakhutdinov]{burda2015importance}
Yuri Burda, Roger Grosse, and Ruslan Salakhutdinov.
\newblock Importance weighted autoencoders.
\newblock \emph{arXiv preprint arXiv:1509.00519}, 2015.

\bibitem[Khemakhem et~al.(2020)Khemakhem, Kingma, Monti, and
  Hyvarinen]{Khemakhem2020}
Ilyes Khemakhem, Diederik Kingma, Ricardo Monti, and Aapo Hyvarinen.
\newblock Variational autoencoders and nonlinear ica: A unifying framework.
\newblock In \emph{International Conference on Artificial Intelligence and
  Statistics}, pages 2207--2217. PMLR, 2020.

\bibitem[Stokes et~al.(2023)Stokes, Shrier, and Steele]{stokes2023simulation}
Tyrel Stokes, Ian Shrier, and Russell Steele.
\newblock Simulation experiments as a causal problem.
\newblock \emph{arXiv preprint arXiv:2308.10823}, 2023.

\bibitem[Singh et~al.(2022)Singh, Mhasawade, and
  Chunara]{singh2022generalizability}
Harvineet Singh, Vishwali Mhasawade, and Rumi Chunara.
\newblock Generalizability challenges of mortality risk prediction models: A
  retrospective analysis on a multi-center database.
\newblock \emph{PLOS Digital Health}, 1\penalty0 (4):\penalty0 e0000023, 2022.

\bibitem[Zimmerman et~al.(2006)Zimmerman, Kramer, McNair, and
  Malila]{zimmerman2006apache}
Jack~E Zimmerman, Andrew~A Kramer, Douglas~S McNair, and Fern~M Malila.
\newblock Acute physiology and chronic health evaluation (apache) iv: hospital
  mortality assessment for today’s critically ill patients.
\newblock \emph{Critical care medicine}, 34\penalty0 (5):\penalty0 1297--1310,
  2006.

\bibitem[Cole and Hern{\'a}n(2008)]{cole2008constructing}
Stephen~R Cole and Miguel~A Hern{\'a}n.
\newblock Constructing inverse probability weights for marginal structural
  models.
\newblock \emph{American journal of epidemiology}, 168\penalty0 (6):\penalty0
  656--664, 2008.

\bibitem[Xiao et~al.(2013)Xiao, Moodie, and Abrahamowicz]{xiao2013comparison}
Yongling Xiao, Erica~EM Moodie, and Michal Abrahamowicz.
\newblock Comparison of approaches to weight truncation for marginal structural
  cox models.
\newblock \emph{Epidemiologic Methods}, 2\penalty0 (1):\penalty0 1--20, 2013.

\bibitem[Kennedy(2023)]{kennedy2023towards}
Edward~H Kennedy.
\newblock Towards optimal doubly robust estimation of heterogeneous causal
  effects.
\newblock \emph{Electronic Journal of Statistics}, 17\penalty0 (2):\penalty0
  3008--3049, 2023.

\bibitem[van~der Laan et~al.(2024)van~der Laan, Carone, and
  Luedtke]{van2024combining}
Lars van~der Laan, Marco Carone, and Alex Luedtke.
\newblock Combining t-learning and dr-learning: a framework for
  oracle-efficient estimation of causal contrasts.
\newblock \emph{arXiv preprint arXiv:2402.01972}, 2024.

\bibitem[Li and Li(2019)]{li2019propensity}
Fan Li and Fan Li.
\newblock Propensity score weighting for causal inference with multiple
  treatments.
\newblock 2019.

\end{thebibliography}

\appendix

\section{Proof Conditional Observable distribution}\label{app:cond proof}

 The proof largely adapts a proof from \cite{zhou2023domain}.\\
 
 We will derive the conditional distribution with respect to the joint distribution. By assumption we have that $P^T(Y|X^u,X^o) = P^S(Y|X^u,X^o)$. The proof procedes in the discrete  case, but can be adapted to continuous distributions with extra care in notation.\\

Let $\mathcal{A} = x^u:x^u\odot m = \tilde{x}^u$, be the set of potentially unobserved variables $X^u$ such that the element wise product with the missingness indicator gives us the observed covariates. Let $d$ be a domain $d \in \{T,S\}$, for Target and Source.

\begin{align*}
P^d(\Y,\Xu = \tilde{x}^u, \M =m, \Xo = x^o) &=
\sum_{\mathcal{A}} P^d(\Y,\Xu = \tilde{x}^u, \M =m, \Xo = x^o, \Xu = x^u)\\
&= \sum_{\mathcal{A}}P^d(\Y, \M =m, \Xo = x^o, \Xu = x^u)\\
&= \sum_{\mathcal{A}}P^d(\M =m|\Y, \Xo = x^o, \Xu = x^u)\times
P^d(\Y, \Xo = x^o, \Xu = x^u)\\
&= \sum_{\mathcal{A}}P^d(\M =m|\Xo = x^o, \Xu = x^u)\times P^d(\Y, \Xo = x^o, \Xu = x^u)
\end{align*}

As shown in \cite{zhou2023domain}, in the special case of MAR, we have that $P^d(\M =m|\Xo = x^o, \Xu = x^u) = P^d(\M =m|\Xo = x^o)$ which can be taken outside of the sum over $\mathcal{A}$. In that case the missingness mechanism drops out of the conditional distribution: i.e the observed conditional distribution does not depend on the missingness mechanism. However, when we have missigness not at random it does not drop out and we have that:

\begin{align*}
&P^d(\Y|\Xo,\txu,\M)\\
&= \frac{P^d(\Y,\Xo,\txu,\M)}{\sum_y P^d(\Y =y,\Xo,\txu,\M)}\\
&= \frac{\sum_{\mathcal{A}}P^d(\M =m|\Xo = x^o, \Xu = x^u)P^d(\Y, \Xo = x^o, \Xu = x^u)}{\sum_y \sum_{\mathcal{A}}P^d(\M =m|\Xo = x^o, \Xu = x^u)P^d(\Y =y, \Xo = x^o, \Xu = x^u)}\\
&= \frac{\sum_{\mathcal{A}}P^d(\M =m|\Xo = x^o, \Xu = x^u)P^d(\Y|\Xo = x^o, \Xu = x^u)P^d(\Xu = x^u|\Xo = x^o)P^d(\Xo = x^o)}{\sum_y \sum_{\mathcal{A}}P^d(\M =m|\Xo = x^o, \Xu = x^u)P^d(\Y =y|\Xo = x^o, \Xu = x^u)P^d(\Xu = x^u|\Xo = x^o)P^d(\Xo = x^o)}\\
&= \frac{\sum_{\mathcal{A}}P^d(\M =m|\Xo = x^o, \Xu = x^u)P^d(\Y|\Xo = x^o, \Xu = x^u)P^d(\Xu = x^u|\Xo = x^o)}{\sum_y \sum_{\mathcal{A}}P^d(\M =m|\Xo = x^o, \Xu = x^u)P^d(\Y =y|\Xo = x^o, \Xu= x^u)P^d(\Xu = x^u|\Xo = x^o)}\\
&= \frac{\sum_{\mathcal{A}}P^d(\M =m|\Xo = x^o, \Xu = x^u)P(\Y|\Xo = x^o, \Xu = x^u)P^d(\Xu = x^u|\Xo = x^o)}{\sum_y \sum_{\mathcal{A}}P^d(\M =m|\Xo = x^o, \Xu = x^u)P(\Y =y|\Xo = x^o, \Xu = x^u)P^d(\Xu = x^u|\Xo = x^o)}
\end{align*}

by assumption we have that $P^T(\Y|\Xo = x^o, \Xu = x^u) = P^S(\Y|\Xo = x^o, \Xu = x^u) = P(\Y|\Xo = x^o, \Xu = x^u)$. \\

However, notice that this depends on both the missingness mechanism $P^d(\M|\Xu,\Xo)$ and the conditional distribution of the potentially unobserved variables and the observed covariates $P^d(\Xu|\Xo)$ which both potentially depend on the domain. This means that different MNAR mechanisms can imply both a Domain and Covariate Shift problem even when the underlying full conditional distributions are the same.

\section{Decomposition Proof}\label{app:decomp proof}

First we show that we can re-express the minimization of the mean-squared error (MSE) in terms of a minimization problem with respect to the best predictor, $\Ept[\Y|\Xo,\txu,\M]$:
\begin{align}
    &\underset{g\in \mathcal{G}}{\arg\min} ~\mathbb{E}_{\mathbb{P}^{T}}[(\Y - g(\Xo,\txu,\M))^2] =\\
    &\underset{g\in \mathcal{G}}{\arg\min}~ \mathbb{E}_{\mathbb{P}^{T}}[(\mathbb{E}_{\mathbb{P}^{T}}[\Y|\Xo,\txu,\M] - g(\Xo,\txu,\M))^2] +
    \mathbb{E}_{\mathbb{P}^{T}}[(\Y - \mathbb{E}_{\mathbb{P}^{T}}[\Y|\Xo,\txu,\M])^2]\\
    &=\underset{g\in \mathcal{G}}{\arg\min} ~\mathbb{E}_{\mathbb{P}^{T}}[(\mathbb{E}_{\mathbb{P}^{T}}[\Y|\Xo,\txu,\M] - g(\Xo,\txu,\M))^2],
\end{align}
where $g(\cdot)$ is an estimator in estimator class $\mathcal{G}$. Then we can re-express this minimization problem in terms of the probability of missingness and the imputation error:

\begin{align*}
&E^T[(E^T[\Y|\Xo,\txu,\M] - \hat{\mu}(\Y|\Xo,\txu = \hat{\X}^u,\M))^2]\\
&= P(\M =0)E^T[(E^T[\Y|\Xo,\txu,\M] - \hat{\mu}(\Y|\Xo,\txu = \hat{\X}^u,\M))^2|\M=0]\\
&+P(\M =1)E^T[(E^T[\Y|\Xo,\txu,\M] - \hat{\mu}(\Y|\Xo,\txu,\M))^2|\M=1]\\
&\pto P(\M =0)E^T[(E^T[\Y|\Xo,\txu,\M=0] - E^T[\Y|\Xo,\txu = \hat{\X}^u,\M=0])^2|\M=0]+\\
&P(\M =1)E^T[(E^T[\Y|\Xo,\txu,\M=1] - \plimn\hat{\mu}(\Y|\Xo,\txu=\hat{\X}^u,\M=1))^2|\M=1]\\
&= P(\M =0)E^T[(E^T[\Y|\Xo,\txu,\M=0] - \plimn\hat{\mu}(\Y|\Xo,\txu= \Xu + (\hat{\X}^u-\Xu),\M=1))^2|\M=0].
\end{align*}

\section{Learning the Conditional Distribution from Complete-Case Data}

In particular, suppose that $\Y\indep \M|\Xo,\Xu$, then the conditional outcome distribution is identifiable from the complete-case data, i.e. in notation
\begin{align*}
    \mathbb{P}(\Y|\M,\Xo,\Xu) &= \mathbb{P}(\Y|\Xo,\Xu)\label{cond1}\\
    &= \mathbb{P}(\Y|\Xo,\Xu, \M = 1)\\ 
    &= \mathbb{P}(\Y|\Xo,\txu, \M = 1), 
\end{align*}
since $\Y\indep \M|\Xo,\Xu$. Notice that $\mathbb{P}(\Y|\Xo,\txu, \M = 1)$ is a function of only observable variables. This means that we can use samples from $\mathbb{P}^S(\Y|\Xo, \txu, \M = 1)$ to estimate both $\mathbb{P}^S(\Y|\M,\Xo,\Xu)$ and $\mathbb{P}^T(\Y|\M,\Xo,\Xu)$ since they are assumed equivalent. 

\section{Reduction to Covariate Shift}

In appendix \ref{app:cond proof} we showed that in the case of MNAR, we are unable to reduce the problem to a covariate shift problem where we can reweight the samples of the source data to recover the loss function in the target directly. However, we can show in the case that we have access to an imputation procedure which is able to recover samples from a distribution which resembles the truth sufficiently accurately then the problem reduces to the usual form, even in the case of covariate shift.

Suppose that we have an imputation estimator, $\hat{\X}^u$ with the property that:

\begin{align}
P(|\hat{\X}^u - \Xu| \geq \epsilon) \leq g(n)
\end{align}
where $\lim_{n\to\infty} g(n) = 0$, then $\hat{\X}^u$ converges to $\Xu$ in probability and thus in distribution. Then we have that:

\begin{align}
\mathbb{P}^S(\Y,\Xo,\hat{\X}^u,\M) &\overset{d}{\to} \mathbb{P}^S(\Y,\Xo,\Xu,\M)\\
&= \mathbb{P}^S(\Y|\Xo,\Xu,\M)\mathbb{P}^S(\M|\Xo,\Xu)\mathbb{P}^S(\Xo,\Xu) \\
&= \mathbb{P}^S(\Y|\Xo,\Xu,\M)\frac{\mathbb{P}^S(\M|\Xo,\Xu)\mathbb{P}^S(\Xo,\Xu)}{\mathbb{P}^T(\M|\Xo,\Xu)\mathbb{P}^T(\Xo,\Xu)} \mathbb{P}^T(\M|\Xo,\Xu)\mathbb{P}^T(\Xo,\Xu)
\end{align}

which allows us to rewrite the minimization problem in the usual way.

\section{Additional Simulation Experiments}
In this section we detail additional simulation results not shown in the main text. The code to reproduce the simulations can be found at the following \href{https://anonymous.4open.science/r/Missingness-shift-MNAR-34D8/README.md}{anonymized github repository}. In the repository we adapt code to fit the GINA and Not-MIWAE algorithms from the original implementations found \href{https://github.com/microsoft/project-azua}{here} and \href{https://github.com/nbip/notMIWAE}{here} respectively.


\begin{figure}
\centering
\scalebox{.45}{
 \includegraphics{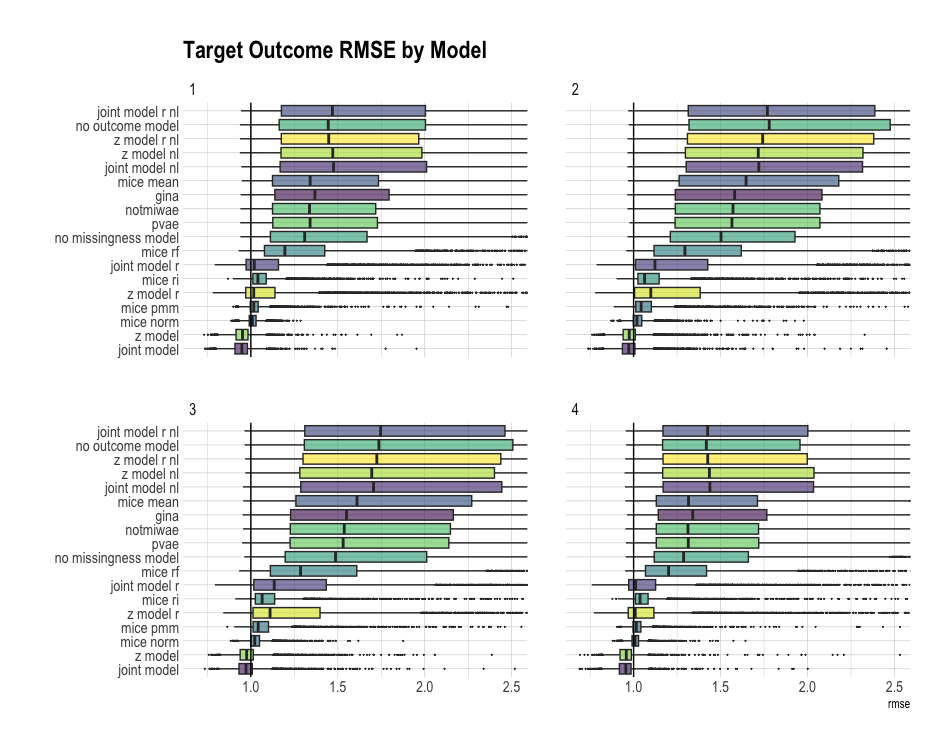}
 }
 \caption{Performance of various imputation models with respect to root mean squared error (RMSE) for target domain
adaptation stratified by data generating DAGs 1 to 4 in the simulation study.}\label{fig:marg-14}
\end{figure}

\begin{figure}
\scalebox{.45}{
 \includegraphics{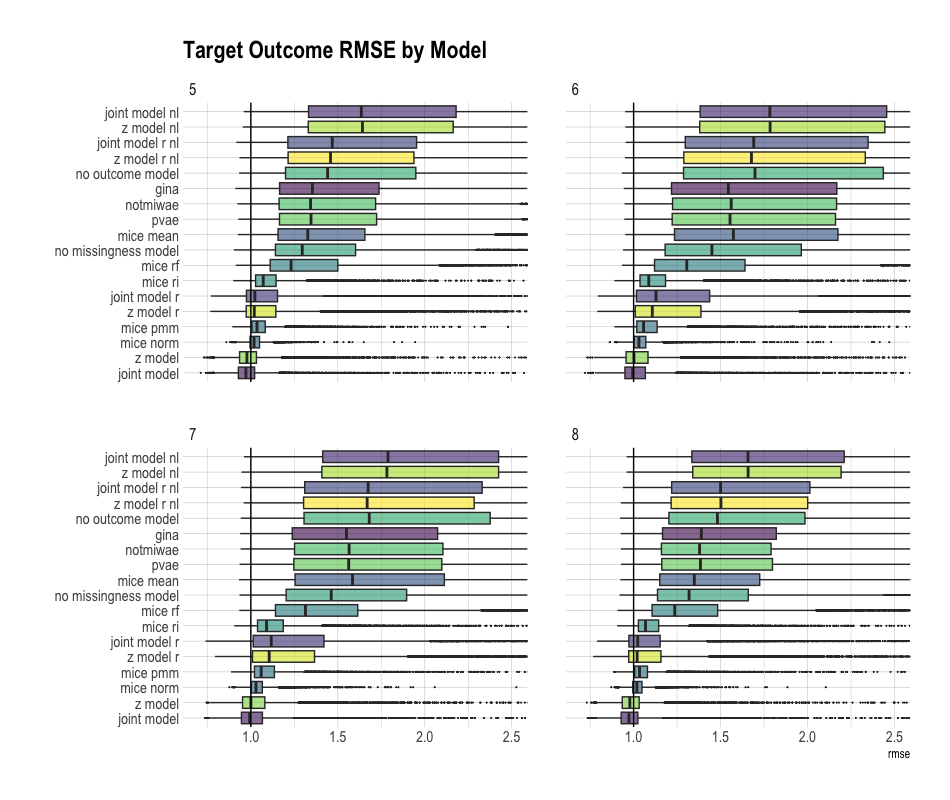}
 }
 \caption{Performance of various imputation models with respect to root mean squared error (RMSE) for target domain
adaptation stratified by data generating DAGs 5 to 8 in the simulation study.}\label{fig:marg-58}
\end{figure}

\begin{figure}[ ]
\centering
\includegraphics[scale = 0.5]{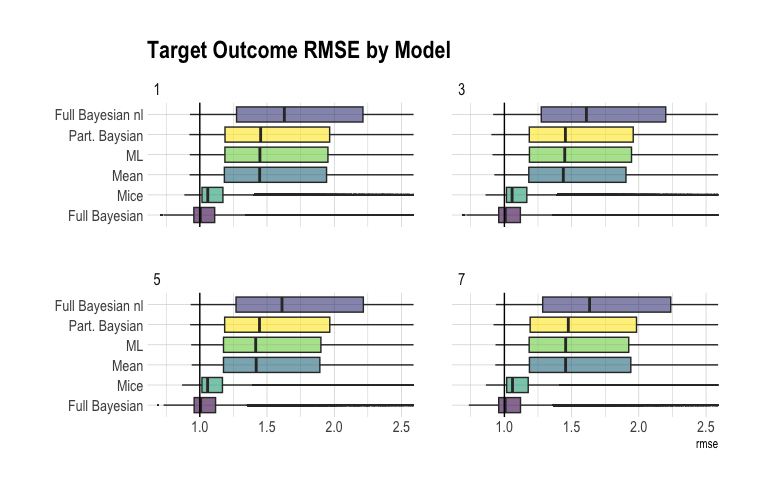}\caption{Performance of various imputation models with respect to root mean squared error (RMSE) for target domain
adaptation stratified by varying level of nonlinearity (in increasing order from 1 to 7) in the relationship between covariates in the simulation study. 1 refers to a linear relationship where as 7 refers to neural network with 7 layers.}\label{fig:marg-cat-nl}
\end{figure}

\end{document}